\documentclass[english,aps,amssymb,nofootinbib,notitlepage]{revtex4-1}
\usepackage{mathptmx}

\usepackage[T1]{fontenc}
\usepackage[utf8]{inputenc}
\usepackage{color}
\usepackage{babel}
\usepackage{float}
\usepackage{units}
\usepackage{textcomp}
\usepackage{amsmath}
\usepackage{amssymb}
\usepackage{wasysym}
\usepackage{graphicx}
\usepackage{esint}
\usepackage[unicode=true,
 bookmarks=false,
 breaklinks=false,pdfborder={0 0 1},backref=section,colorlinks=false]
 {hyperref}
\hypersetup{pdftitle={Fitting and forecasting non-linear coupled dark energy},
 pdfauthor={Santiago Casas},
 pdfsubject={Cosmology},
 pdfkeywords={Cosmology}}

\makeatletter

\newcommand{\noun}[1]{\textsc{#1}}
\providecommand{\tabularnewline}{\\}

\@ifundefined{textcolor}{}
{%
 \definecolor{BLACK}{gray}{0}
 \definecolor{WHITE}{gray}{1}
 \definecolor{RED}{rgb}{1,0,0}
 \definecolor{GREEN}{rgb}{0,1,0}
 \definecolor{BLUE}{rgb}{0,0,1}
 \definecolor{CYAN}{cmyk}{1,0,0,0}
 \definecolor{MAGENTA}{cmyk}{0,1,0,0}
 \definecolor{YELLOW}{cmyk}{0,0,1,0}
}

\usepackage{babel}
\usepackage{babel}
\usepackage{babel}
\usepackage{babel}
\usepackage{babel}
\usepackage{babel}

\definecolor{green}{rgb}{0.09, 0.45, 0.27}

\makeatother

\begin{document}

\title{Fitting and forecasting coupled dark energy in the non-linear regime}

\author{Santiago Casas$^{1}$, Luca Amendola$^{1}$, Marco Baldi$^{2,3,4}$,
Valeria Pettorino$^{1}$, Adrian Vollmer$^{1}$}

\affiliation{$^{1}$Institut f{ü}r Theoretische Physik, Ruprecht-Karls-Universit{ä}t
Heidelberg, Philosophenweg 16, 69120 Heidelberg, Germany \\
 $^{2}$Dipartimento di Fisica e Astronomia, Alma Mater Studiorum
Università di Bologna, viale Berti Pichat, 6/2, I-40127 Bologna, Italy;
\\
 $^{3}$INAF - Osservatorio Astronomico di Bologna, via Ranzani 1,
I-40127 Bologna, Italy; \\
 $^{4}$INFN - Sezione di Bologna, viale Berti Pichat 6/2, I-40127
Bologna, Italy. }
\begin{abstract}
We consider cosmological models in which dark matter feels a fifth
force mediated by the dark energy scalar field, also known as coupled
dark energy. Our interest resides in estimating forecasts for future
surveys like Euclid when we take into account non-linear effects,
relying on new fitting functions that reproduce the non-linear matter
power spectrum obtained from N-body simulations.

We obtain fitting functions for models in which the dark matter-dark
energy coupling is constant. Their validity is demonstrated for all
available simulations in the redshift range $z=0-1.6$ and wave modes
below $k=\unit[10]{h/Mpc}$. These fitting formulas can be used to
test the predictions of the model in the non-linear regime without
the need for additional computing-intensive N-body simulations. We
then use these fitting functions to perform forecasts on the constraining
power that future galaxy-redshift surveys like Euclid will have on
the coupling parameter, using the Fisher matrix method for galaxy
clustering (GC) and weak lensing (WL). We find that by using information
in the non-linear power spectrum, and combining the GC and WL probes,
we can constrain the dark matter-dark energy coupling constant squared,
$\beta^{2}$, with precision smaller than 4\% and all other cosmological
parameters better than 1\%, which is a considerable improvement of
more than an order of magnitude compared to corresponding linear power
spectrum forecasts with the same survey specifications. 
\end{abstract}
\maketitle

\section{Introduction}

\global\long\def\lcdm{\Lambda\textrm{CDM}}
 \global\long\def\jcap{JCAP}

Cosmological data have reached a level of precision that forces us
to address effects which may have been safely neglected in the past.
Among them, predicting the behaviour of theoretical models at non-linear
scales gives a more concrete chance to disentangle $\Lambda$CDM from
alternative scenarios. The non-linear power spectrum contains useful
information encoded in the clustering of galaxies and the weak lensing
correlations, which can help to constrain more tightly the cosmological
parameters and remove certain degeneracies between them. The drawback
relies in the fact that the full estimate of the non-linear power
spectrum requires time-demanding and computationally expensive N-body
simulations. As a consequence, there exist simulations only for a
handful of models beyond $\Lambda$CDM on a very limited subset of
the parameter space of dark energy theories. This makes it hard to
use the non-linear power spectrum in forecasts (see e.g. \citep{2011JCAP...03..026C}),
both when doing cosmological parameter estimation via Monte Carlo
simulations and with the simpler approximation of the Fisher matrix
formalism. The alternative is to employ fitting functions for the
non-linear power spectrum, like Halofit \citep{smith_stable_2003,takahashi_revising_2012}
or machine learning estimators like the cosmological emulator (CosmicEmu)
\citep{heitmann_coyote_2010,lawrence_coyote_2010,heitmann_coyote_2014}
and the interpolator PkANN \cite{agarwal_pkann_2012,agarwal_pkann_2014}
based on artificial neural networks. All these methods are calibrated
or trained using large sets of N-body simulations and typically permit
an exploration of the parameter space just around a fiducial $\lcdm$
cosmology or $w\mbox{CDM }$cosmology with a constant $w$. An extension
of Halofit to account for massive neutrinos was realised in \citep{bird_massive_2011},
which was then used in \citep{audren_neutrino_2013} to perform a
parameter estimation using a Markov Chain Monte Carlo technique. Other
authors have recently used polynomial fits to take into account systematic
effects in the non-linear regime like baryonic effects such as in
\cite{bielefeld_cosmological_2014} or to parametrize the cosmological
dependence of non-linear clustering, beyond the Zeldovich approximation
(see \cite{mohammed_analytic_2014}). These fitting functions have,
however, an intrinsic error that usually increases when scales become
highly non-linear; therefore one has to be aware of the range in scales
and redshifts they were designed to work on, in order to keep the
errors induced on the power spectrum under control.

As an alternative to fitting formulae and N-body simulations, there
has been in recent years a substantial progress in semi-analytical
methods to calculate the non-linear power spectrum at mildly non-linear
scales, such as for example renormalized perturbation theory \citep{crocce_renormalized_2006}
and all other resummation methods derived from it (for example \cite{Anselmi2012}),
the time renormalisation group (TRG) \citep{Pietroni_2008} and effective
field theories of large scale structure \cite{baumann_cosmological_2010}.
Most of these approaches claim to reach percent accuracy at the baryon
acoustic oscillation (BAO) scale and a bit beyond the second BAO peak,
but they are still not able to predict what happens at highly non-linear
scales, when the single stream approximation does not hold any longer.

In this paper, we will use N-Body simulations to find fitting functions
for a class of models beyond $\Lambda$CDM usually refered to as coupled
Dark Energy (CDE). These models, widely discussed in literature \citep{Wetterich_1995, Amendola_2000,Amendola_2004,pettorino_baccigalupi_2008},
involve an extra degree of freedom, associated to a scalar field that
provides acceleration and mediates a fifth force, in addition to gravity,
which is felt by dark matter particles only. Semi-analytical non-linear
analysis \cite{wintergerst_clarifying_2010,Saracco_etal_2010} and
cosmological N-body simulations within coupled Dark Energy have been
performed by many different groups \citep{baldi_etal_2010,Li_Barrow_2011,maccio_coupled_2004,carlesi_hydrodynamical_2014-1,baldi_codecs_2012,2015arXiv150407243P}
and their effects on large scale structure formation have been identified
and characterised. The power spectrum, halo mass functions and concentration,
halo spin and sphericity, voids and amount of substructures show noticeable
differences compared to a simple $\Lambda$CDM model (see for example
\citep{mainini_mass_2006,maccio_concentration_2008,baldi_clarifying_2011,cui_halo_2012,baldi_effect_2011,sutter_observability_2014}
and the review article \citep{Baldi_2012b}). For a constant coupling,
constraints have been found for a variety of probes \citep{pettorino_constraints_2012,pettorino_testing_2013,amendola_skewness_2004,Xia_2009},
with the latest ones discussed by the Planck collaboration in \citep{planckcollaboration_planck2015_2015}.
Recently there have been attempts to constrain more general couplings
between dark matter (DM) and dark energy (DE) using large-scale structure
\citep{creminelli_single-field_2014}, CMB \citep{morris_cosmicmicrowave_2014}
or laboratory experiments \citep{hamilton_atominterferometry_2015}.
Forecasts on coupled dark energy using galaxy clustering (GC) and
weak lensing (WL) measurements for future surveys like Euclid have
been discussed in \citep{amendola2012testing}, but have been performed
using only linear power spectra for the CDE models. The TRG method
has been extended to coupled Dark Energy \citep{saracco_non-linear_2010}
and to massive neutrinos \citep{lesgourgues_non-linear_2009}. However
the TRG method does not produce a reliable estimation of the power
spectrum for scales larger than $k\approx0.3h/\mbox{Mpc}$, which
makes them less suitable for forecasts which attempt to extract information
on highly non-linear scales.

The CoDECS (Coupled Dark Energy Cosmological Simulations) set of N-body
simulations \citep{baldi_codecs_2012} has shown that CDE models have
characteristic and measurable features in the morphology and history
of non-linear structures, such as halos, subhalos and voids,\textcolor{black}{{}
and therefore in the non-linear power spectrum.}

The aim of this paper is to create fitting functions which are valid
in the observable regime of non-linear perturbations at all interesting
redshifts and reproduce the subtle effects of coupled dark energy
on the non-linear power spectrum while allowing us to vary the different
parameters of the model. We use this to perform forecasts of cosmological
parameters assuming coupled dark energy as the fiducial model, using
galaxy clustering and weak lensing as observational tools, as expected
for future surveys like Euclid \cite{laureijs_euclid_2011,amendola_cosmology_2012-short}.
We do a careful treatment of errors and systematics, so that we take
into account all errors induced by our fitting functions, the cosmic
emulators and the extraction of the power spectrum from the N-body
simulation into the analysis, systematics related to non-linear effects
that affect the redshift space distortions and the lensing signals.
In this way we obtain a conservative estimate on how well a probe
like Euclid, will be able to measure a DM-DE coupling.

\section{Coupled Dark Energy\label{sec:Coupled-Dark-Energy}}

In this section we briefly review the main equations governing the
coupled DE models that will be investigated in the present work. For
a more detailed and extended discussion and for the full derivation
of the equations we refer the interested reader to some of the original
publications on the subject \citep{Amendola_2000,Amendola_2004,pettorino_baccigalupi_2008,baldi_etal_2010}.

The background evolution for the coupled DE scenario model is described
by the following equations, in which the subscripts $r$, $b$, $c$
and $\phi$, indicate radiation, baryons, cold dark matter (CDM) and
the dark energy scalar field, respectively:

\begin{eqnarray}
\ddot{\phi}+3H\dot{\phi}+\frac{dV}{d\phi} & = & \sqrt{\frac{2}{3}}\beta(\phi)\frac{\rho_{c}}{M_{Pl}}\,\,,\label{eq:quint-kleingordon}\\
\dot{\rho}_{c}+3H\rho_{c} & = & -\sqrt{\frac{2}{3}}\beta(\phi)\frac{\rho_{c}\dot{\phi}}{M_{Pl}}\,\,,\label{eq:cdm-back-density}\\
\dot{\rho}_{b}+3H\rho_{b} & = & 0\,\,,\\
\dot{\rho}_{r}+4H\rho_{r} & = & 0\,\,,\\
3H^{2} & = & \frac{1}{M_{Pl}^{2}}(\rho_{b}+\rho_{c}+\rho_{r}+\rho_{\phi})\,\,.
\end{eqnarray}

We express from now on the scalar field $\phi$ in units of the Planck
mass $M_{pl}\equiv1/\sqrt{8\pi G}$, and choose as potential $V(\phi)$
an exponential $V(\phi)=Ae^{-\alpha\phi}$ \citep{Lucchin_Matarrese_1984,Wetterich_1988}.
The coupling function $\beta(\phi)$ defines the strength of the interaction
between the DE fluid and CDM particles and in the present work we
will restrict our analysis to the simplified case of a constant coupling
$\beta(\phi)=\beta$, although in general it could be a field-dependent
quantity \cite{Amendola_2004,Baldi_2011a}.

The current constraints on a coupling to ordinary matter are very
tight. The ``post-Einstein'' coupling parameter $\bar{\gamma}$
that measures the local admixture of a spin-0 field to gravity is
constrained in Solar System experiments (see e.g. the PDG review \cite{Agashe:2014kda}
and also \citep{Will_2005,Bertotti_Iess_Tortora_2003}) roughly to
$|\bar{\gamma}|\le4\cdot10^{-5}$ . This parameter enters the modification
of the effective Newton constant as $G_{eff}=G_{N}(1-\bar{\gamma}/2)$;
in our notation, this is $G_{eff}=G_{N}(1+4\beta^{2}/3)$ (see below)
and therefore $\beta^{2}=-3\bar{\gamma}/8$. A coupling $\beta_{baryons}^{2}$
appears then constrained to be smaller than $10^{-5}$ roughly, and
we assume therefore that is completely negligible. As a consequence,
baryons follow the usual geodesics of a FLRW cosmology, which allows
coupled DE to pass the stringent local gravity constraints without
the need to employ any screening mechanism \citep{2015arXiv150203888H}.

Due to the exchange of energy between DE and CDM, the energy density
of the latter will no longer scale as the cosmic volume, and by assuming
the conservation of the CDM particle number one can derive the time
evolution of the CDM particle mass by integrating Eq.(\ref{eq:cdm-back-density})
between the present time ($z=0$) and any other redshift $z$:

\begin{equation}
m_{c}(z)=m_{c,0}e^{-\beta(\phi(z)-\phi(0))}\,\,.
\end{equation}

\begin{table}
\centering{}\label{tab:models} %
\begin{tabular}{ccc|c}
\textbf{Parameter}  & \textbf{Explanation}  & \textbf{Value}  & \textbf{Reference $\Lambda$CDM}\tabularnewline
\hline 
\hline 
$A$  & Potential normalization  & $0.0218$  & -- \tabularnewline
$\alpha$  & Potential slope  & $0.08$  & -- \tabularnewline
$\phi(z=0)$  & Scalar field normalization  & $0$  & -- \tabularnewline
$\beta$  & Coupling parameter  & $\{0.05,\,0.10,\,0.15\}$  & -- \tabularnewline
$w_{\phi}(z=0)$  & DE equation of state  & $\{-0.997,\,-0.995,\,-0.992\}$  & $-1$ \tabularnewline
$\sigma_{8}(z=0)$  & Power spectrum amplitude  & $\left\{ 0.825,\,0.875,\,0.967\right\} $  & $0.809$ \tabularnewline
\hline 
\end{tabular}\protect\protect\caption{\label{tab:normalization-conventions-codecs} The main parameters
and normalizations of the three coupled DE models of the \noun{\small{}{}CoDECS}{\small{}
suite considered in the present work. The last column displays the
corresponding values for the reference $\Lambda$CDM cosmology.}}
\end{table}

At the level of linear perturbations, coupled DE models are characterised
by a different evolution of the baryonic and CDM density fluctuations,
as a consequence of the selective interaction between DE and CDM particles
only. In the sub-horizon limit, for which $aH/k\ll1$, linear perturbations
in coupled dark energy follow the equations \citep{Amendola_2004,pettorino_baccigalupi_2008}:
\begin{eqnarray}
\ddot{\delta}_{c} & = & -2H\left[1-\beta\frac{\dot{\phi}}{H\sqrt{6}}\right]\dot{\delta}_{c}+4\pi G\left[\rho_{b}\delta_{b}+\rho_{c}\delta_{c}\Gamma_{c}\right]\label{linear_cdm}\\
\ddot{\delta}_{b} & = & -2H\dot{\delta}_{b}+4\pi G\left[\rho_{b}\delta_{b}+\rho_{c}\delta_{c}\right]\label{linear_baryons}
\end{eqnarray}
where $\Gamma_{c}\equiv1+\frac{4}{3}\beta^{2}$ represents the effective
``fifth force'' acting on the CDM particles. The term proportional
to $\beta\dot{\phi}$ in equation (\ref{linear_cdm}) is a velocity-dependent
term that modifies the standard cosmological friction; this arises
as a consequence of momentum conservation for the CDM particles and
has a considerable effect on structure formation \citep{baldi_etal_2010,baldi_clarifying_2011,Li_Barrow_2011}.
Since baryons are uncoupled, their perturbations evolve according
to the standard equation. Nonetheless, baryons will still be indirectly
affected by the coupling as the source term on the right-hand side
of equation (\ref{linear_baryons}) includes the CDM density perturbations.

At the level of non-linear perturbations, several methods have been
devised to predict the small scale effects of coupled DE, from semi-analytical
methods like spherical collapse \cite{Pace_Waizmann_Bartelmann_2010},
to time renormalisation group \citep{Saracco_etal_2010} to full N-body
simulations \citep{Maccio_etal_2004,baldi_etal_2010,Li_Barrow_2011,Carlesi_etal_2014a}.
In the following, we will work with the publicly available data of
the \noun{\small{}{}CoDECS}{\small{} simulations \citep{baldi_codecs_2012}
that represents the largest set of cosmological N-body simulations
for coupled DE models to date.}{\small \par}

\section{The CoDECS simulations}

The \noun{\small{}{}CoDECS}{\small{} suite}%
\footnote{see also the public {\small{}{}{}{}{}{}{}{}{}CoDECS} database
at www.marcobaldi.it/CoDECS/%
}{\small{} includes simulated periodic volumes of the universe at different
scales and with different physical ingredients (as e.g. simulations
with and without hydrodynamics) in the context of a series of coupled
DE cosmologies characterised by various choices of the self-interaction
potential $V(\phi)$ and of the coupling functions $\beta(\phi)$.
The simulations have been performed with a suitably modified version
(see \citep{baldi_codecs_2012} for more details on the numerical
implementation) of the widely-used TreePM N-body code }\noun{\small{}{}GADGET}{\small{}
\citep{springel_cosmological_2005}. Such modified version includes
all the relevant effects that characterise coupled DE cosmologies,
from the modified background evolution to the CDM particle mass variation,
the ``fifth-force\textquotedbl{} and the velocity-dependent acceleration
appearing in equation (\ref{linear_cdm}).}{\small \par}

As already stated above, in this work we will consider -- besides
the reference $\Lambda$CDM simulation -- the subset of \noun{\small{}{}CoDECS}{\small{}
runs characterised by an exponential potential and by a constant coupling
function. This consists of three different coupled DE models with
the same potential slope $\alpha$ and with three values of the coupling
$\beta=\left\{ 0.05\,,0,1\,,0.15\right\} $. The short names for these
simulations are respectively EXP001, EXP002 and EXP003. All the models
have been built in order to have the same cosmological parameters
at $z=0$ consistent with the WMAP7 results \citep{komatsu_seven-year_2010},
see Table \ref{tab:ParametersWMAP7}, with the obvious exception of
the value of the equation of state parameter $w_{0}$, that changes
from model to model due to the different dynamics of the DE scalar
field. The present observational constraints on the cosmological parameters
have only slightly changed with the latest updated release of Planck
data \citep{Planck:2015xua} with respect to the assumed WMAP7 values,
and are still good enough for the purposes of this paper, being in
good agreement with large scale structure observations. For what concerns
linear perturbations, all cosmologies have been normalised to have
the same statistics (i.e. the same power spectrum shape and amplitude)
of density fluctuations at the redshift of the Cosmic Microwave Background
$z_{{\rm CMB}}\approx1100$. As a consequence of this choice and of
the different growth associated with the various coupling values,
all the models will have a different normalisation $\sigma_{8}$ of
the linear perturbations amplitude at $z=0$. The main features of
these models are summarised in Table \ref{tab:models}. We refer to
\citep{baldi_codecs_2012} for further details.}{\small \par}

For the purposes of the present work, we will employ the matter power
spectra extracted from the CoDECS runs of these cosmologies at different
redshifts in order to find a fitting formula that captures with high
accuracy the deviations of the coupled DE nonlinear power spectra
from the reference $\Lambda$CDM case.

\begin{table}
\centering{}%
\begin{tabular}{cc}
\textbf{Parameter}  & \textbf{Value} \tabularnewline
\hline 
$H_{0}$  & $70.3\,\mbox{km}\mbox{s}^{-1}\mbox{Mpc}^{-1}$ \tabularnewline
$\Omega_{CDM}$  & $0.226$ \tabularnewline
$\Omega_{DE}$  & $0.729$ \tabularnewline
$\mathcal{A}_{s}$  & $2.42\times10^{-9}$ \tabularnewline
$\Omega_{b}$  & $0.0451$ \tabularnewline
$n_{s}$  & $0.966$ \tabularnewline
\hline 
\end{tabular}\protect\protect\caption{ \label{tab:ParametersWMAP7}The set of cosmological parameters used
in all CoDECS simulations, consistent with the WMAP7 results.}
\end{table}

\section{\label{sec:Folding-method}Extracting the power spectrum at small
scales}

The power spectrum $P(k)$, is defined as the ensemble average of
the density contrast in Fourier space $\left\langle \delta(k)\delta(k')\right\rangle \equiv(2\pi)^{3}P(k)\delta_{D}(k+k')$.
When trying to extract this quantity from an N-body simulation, one
has to take into account several technicalities related to \emph{sampling}
effects which appear as a consequence of treating a \emph{discrete
}distribution of particles. For more details related to this problem
and different solution methods, see \citep{cui_ideal_2008}.

At large scales (or equivalently small $k$) the power spectrum suffers
from uncertainties due to the finite size of the simulation box, since
there are only few independent modes to sample the signal from. On
the other hand, at high $k$, one is limited by the resolution of
the simulation, since one cannot sample wave modes smaller than the
typical grid size $L/N$ where $L$ is the length of one side of the
simulation box and $N$ is the number of particles. This maximum frequency
is called the Nyquist frequency $k_{Ny}\equiv2\pi N/L$ and modes
smaller than this \textcolor{black}{cannot be reliably measured} (this
corresponds to the so-called aliasing effect). For the CoDECS simulations
used in this work, the Nyquist frequency has the value $k_{Ny}\approx2.2$
$\mbox{h}/\mbox{Mpc}$ at present time.

The power spectrum computation embedded in \noun{\small{}{}GADGET-3}{\small{}
that was adopted in the }\noun{\small{}{}CoDECS}{\small{} simulations
employs the so-called folding method developed by \cite{colombi_accurate_2009}
-- which is based on \cite{jenkins_galaxy_1999} -- to calculate the
matter power spectrum for smaller scales than the Nyquist frequency.
Following this method one ends up with two separate power spectra,
$P_{top}$ which is calculated using the simulation particle mesh
(PM) at $k\lesssim k_{Ny}$ and $P_{fold}$ which is the folded power
valid for $k\gtrsim k_{Ny}$. In order to provide a single sampling
of the power spectrum across $k_{Ny}$ the {}{}CoDECS} project employed
a simple interpolation procedure around $k_{Ny}$ by averaging the
two power spectra in the region of overlap. However, this might introduce
some spurious features that appear only when considering the ratio
of two power spectra $P(k)$ at highly non-linear scales. Although
such features are harmless for most practical purposes, for the aims
of the present work it is very important to have accurate ratios of
power spectra, since we want to calculate fitting functions that can
capture the effect of the dark energy - dark matter coupling compared
to $\lcdm$ and therefore we need to correct for these spurious effects.

We then developed a \noun{Python} code that finds the optimal interpolation
and matching between the $P_{top}$ and $P_{fold}$ power spectra.
By evaluating the first and second derivatives of the ratios and minimizing
abrupt changes, it finds the optimal number of points to be removed
from $P_{top}$ due to the aliasing error and the number of points
to be removed from the $P_{fold}$ due to the low sampling error;
at the same time it looks for the optimal linear interpolation weights
between them. Moreover, it cuts off the power spectrum when the shot
noise error ($P_{shot}=1/N$) reaches 10\% of the estimated power
spectrum. This method improves considerably the convergence of the
fitting functions at non-linear scales, allowing us to reach our accuracy
goal of 1-2\% (see section \ref{sec:Fitting-functions} on fitting
functions). In figure (\ref{fig:Error-sources}), the uncertainties
on $P(k)$ are plotted, and the shaded region represents the error
on the power spectrum due to the finite number of modes available.
The clear jump present in this shaded region occurs at the scale in
which the folded and top level power spectra have been matched together,
which corresponds roughly to $k_{Ny}$.

\section{Fitting functions\label{sec:Fitting-functions}}

\begin{figure}
\begin{centering}
\includegraphics[width=0.5\columnwidth]{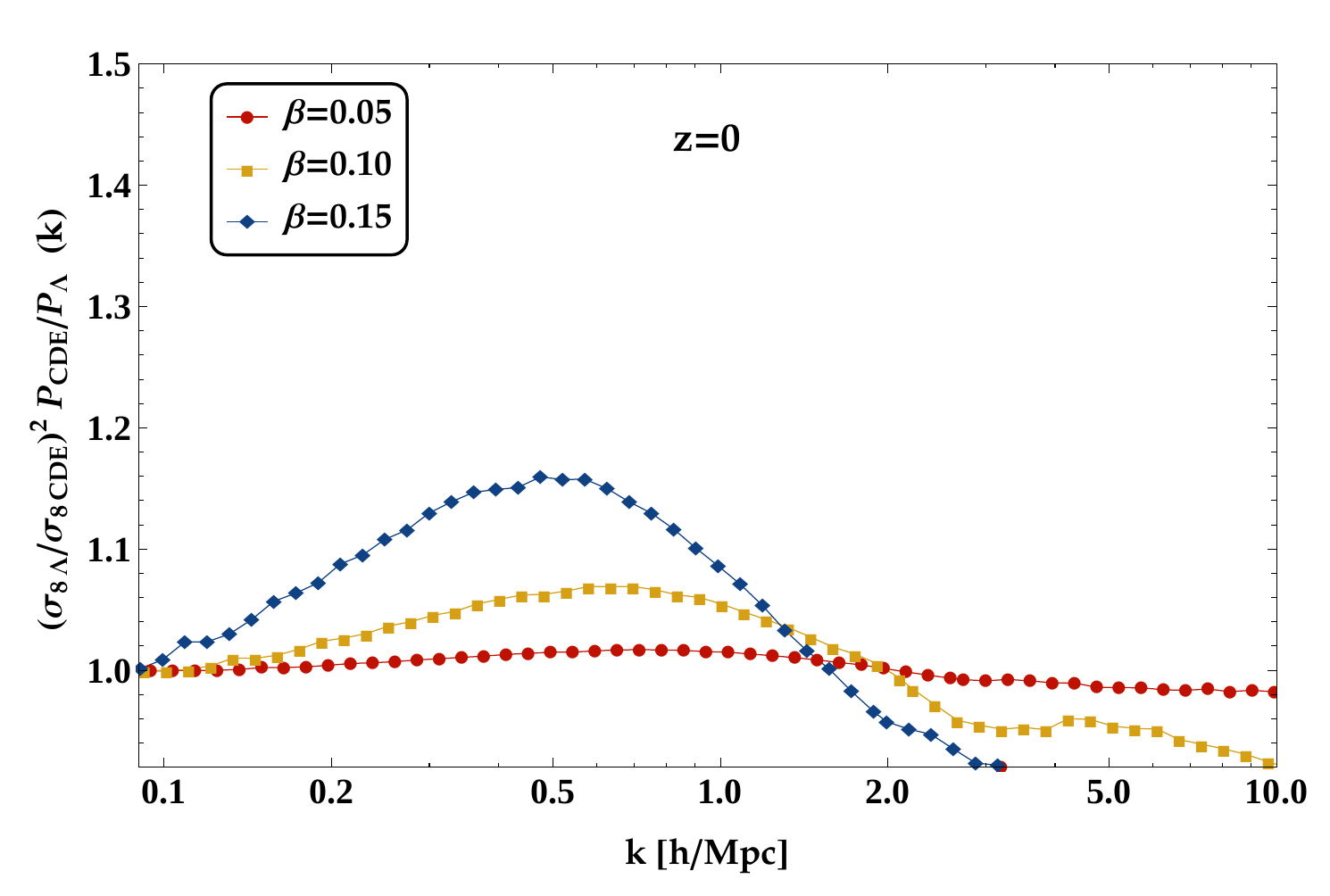}\includegraphics[width=0.5\columnwidth]{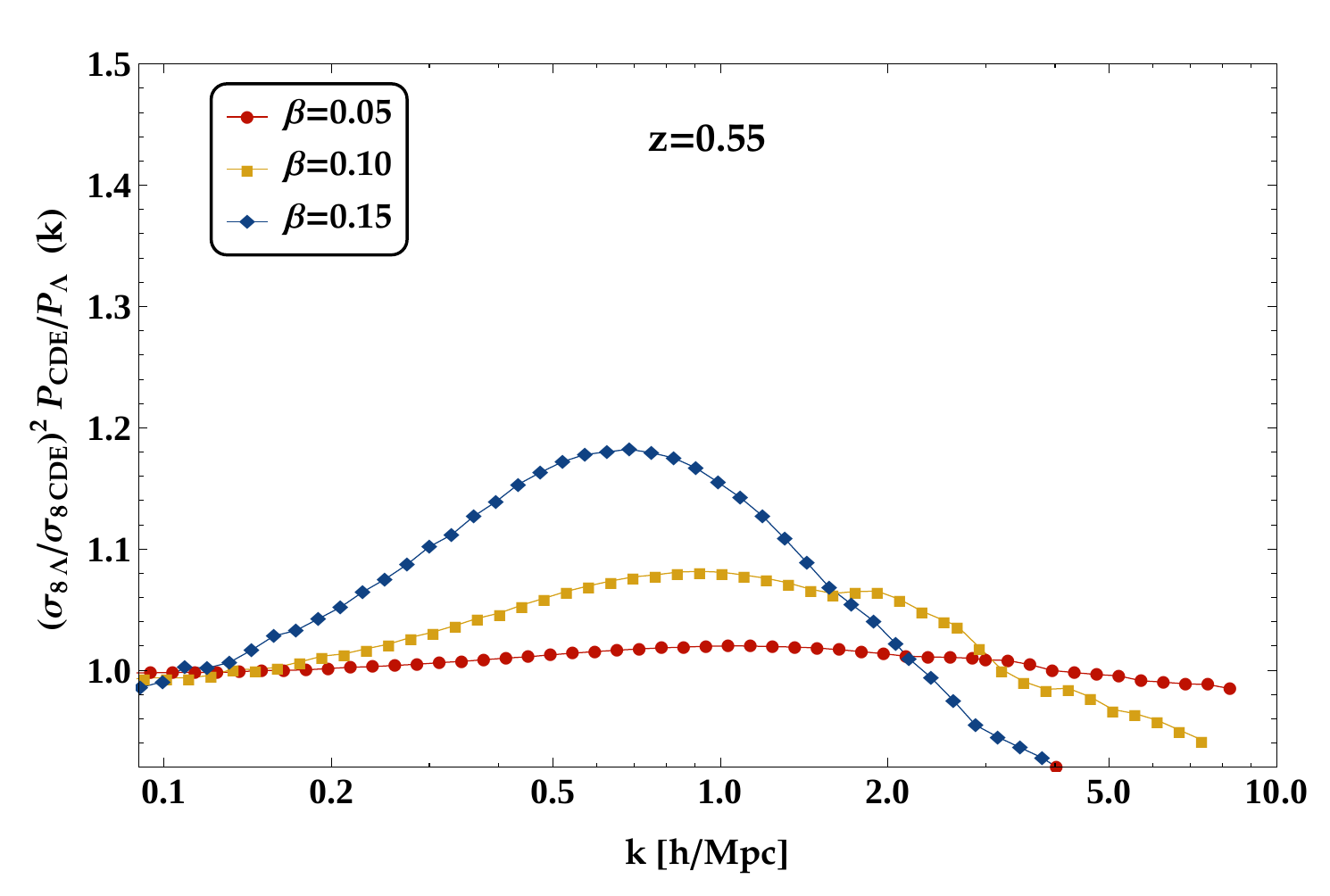}\\
 \includegraphics[width=0.5\columnwidth]{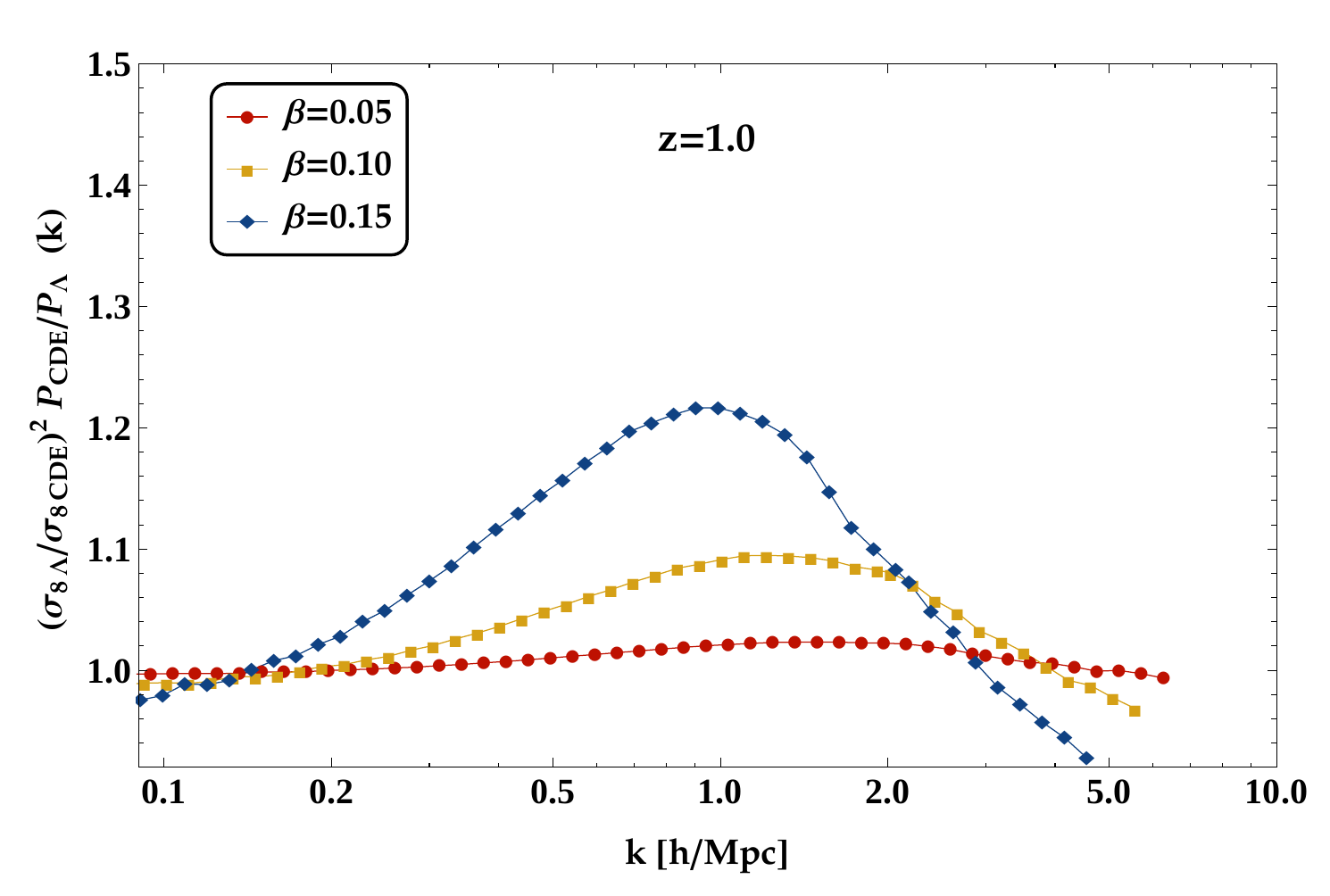}\includegraphics[width=0.5\columnwidth]{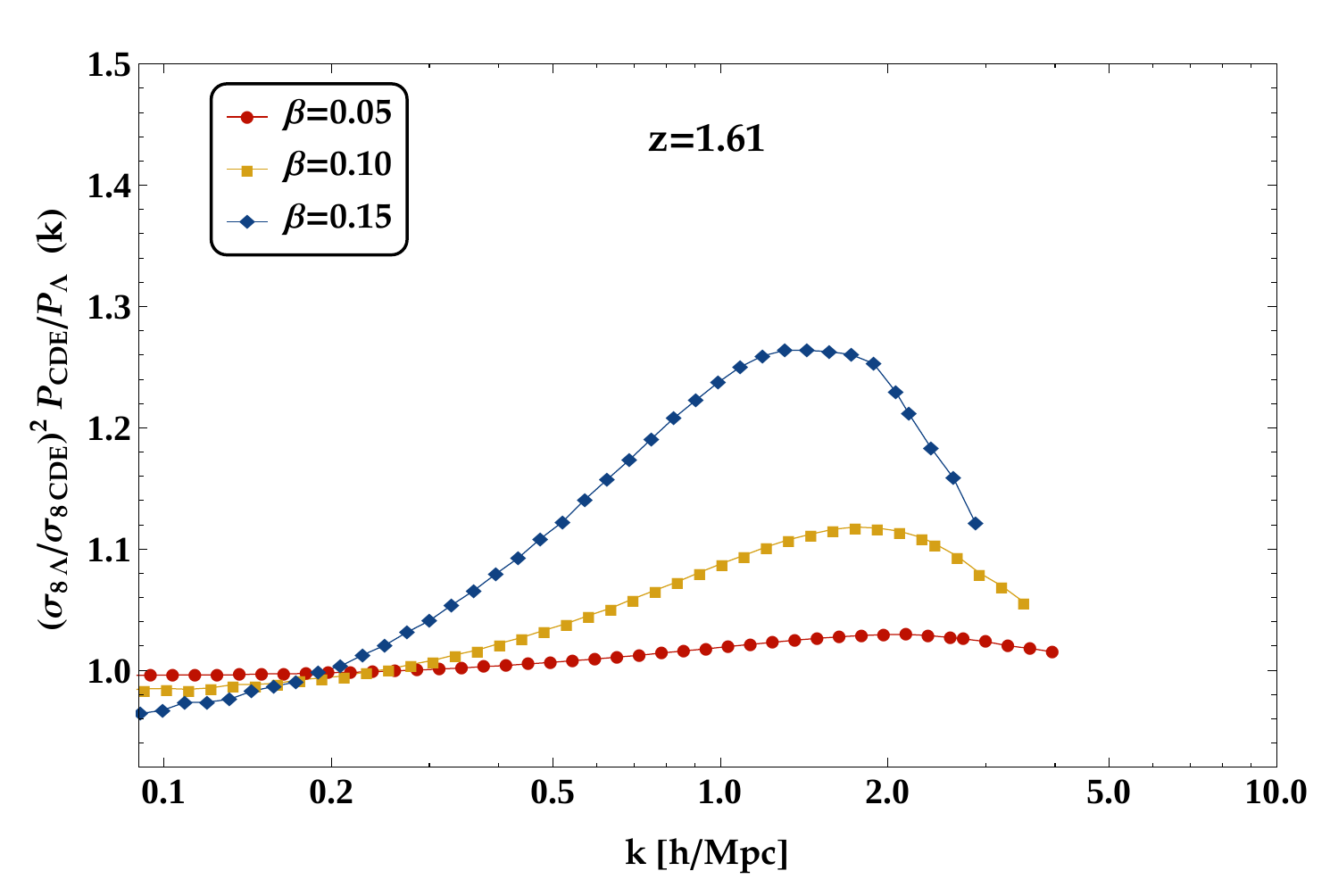} 
\par\end{centering}

\protect\protect\protect\protect\caption{Ratios of the non-linear matter power spectra of the \noun{CoDECS}
CDE models with an exponential potential, with respect to the \noun{CoDECS}
$\protect\lcdm$ power spectra, normalized by their respective $\sigma_{8}^{2}$,
evaluated at four different redshifts and three different coupling
constants $\beta$. Upper left panel: $z=0$, upper right panel: $z=0.55$,
lower left panel: $z=1.0$, lower right panel: $z=1.61$. The blue
line represents the model with strongest coupling \textbf{$\beta=0.15$},
the yellow line $\beta=0.10$ and the red line the smallest available
coupling $\beta=0.05$. In order to be able to observe the net effect
of the $\beta$ coupling at non-linear scales we have divided each
power spectrum by its respective $\sigma_{8}^{2}$.}

\label{fig:Ratio-of-CoDECS} 
\end{figure}

\subsection{The net effect of the DM-DE coupling at non-linear scales}

We now model the net effect of a coupled DE model with an exponential
potential and a constant coupling on top of a fiducial $\Lambda$CDM
non-linear power spectrum, by evaluating the ratio of the non-linear
power spectrum of the coupled DE model, with respect to the one in
the $\Lambda$CDM model, both extracted from the CoDECS simulations.
In figure (\ref{fig:Ratio-of-CoDECS}) we show the ratio $R(k;\beta,z)\equiv P_{Exp}(k;\beta,z)/P_{LCDM}(k;\beta,z)$
as a function of the scale k, for four different redshifts $z=\{0,\,0.55,\,1.0,\,1.61\}$;
each panel contains the curves corresponding to the three available
constant couplings in the CoDECS simulations: \textbf{$\beta=\{0.05,\,0.10,\,0.15\}$}.
Since all coupled DE models have the same amplitude of perturbations
at recombination, an increasing coupling has the effect of inducing
a higher linear normalization $\sigma_{8}$ of the power spectrum
at $z=0$ \citep{baldi_codecs_2012}. Therefore, in order to see the
net effect of the coupling at non-linear scales, each power spectrum
ratio has been re-normalized by dividing each model by its respective
$\sigma_{8}^{2}$, so that at linear scales $k\apprle0.1\mbox{h}/\mbox{Mpc}$
all the ratios are unity. The net effect of the fifth force is a ``bump''
at non-linear scales, whose amplitude increases with higher couplings
and whose maximum is shifted into higher wavenumbers $k$ for higher
redshifts. This extra information imprinted into the non-linear power
spectrum is what we want to use to improve the estimation of parameters
using future surveys. To achieve this, we will fit these curves which
are functions of redshift, physical scale and coupling, using the
minimal number of numerical parameters possible, while keeping the
accuracy goal at the 1\% level.

\subsection{Generating the fitting functions}

The fact that one can observe a clear trend that relates the amplitude
of the signal to an increase of the coupling, together with a shift
of the peak towards larger length scales when looking at smaller redshifts,
is an indication that we should be able to find a relatively simple
fitting formula describing this behaviour, which will be then a function
of $z$, $k$ and the coupling constant $\beta$ only.

To perform the fit, we use a least-squares-minimizing technique, using
the conjugate gradient method \cite{weisstein_least_????}. Taking
into account the particular form in $k$-space of the ratios $R(k;\beta,z)\equiv P_{EXP}(k;\beta,z)/P_{LCDM}(k;\beta,z)$
that we need to fit, we use as an Ansatz different sigmoid functions
to reproduce the particular form of the peak. For each fitting model,
we keep the same functional form for the $k$-dependence of $R(k;\beta,z)$
at all redshifts and for all couplings. We tried 7 different sigmoid
functions as fitting models, but we only show the best two models
M2 and M7 in table \ref{tab:Fitting-models}. All models contain 5
coefficients, which are dependent on the coupling $\beta$ and the
redshift $z$: ($a_{i}=a_{0},\, a_{1},\, c,\, b,\, k_{0}$, with $i=0,...,5$).
The coefficients $k_{0}$ and $b$ determine qualitatively the form
of the peak to be fitted, while the others control mostly the shifting
and the flattening of the peak.

Each coefficient $a_{i}$ is then fitted using a polynomial in $\beta$
and $z$, up to a maximum of third order in powers of $\beta$ and
$z$. Polynomials of order 4 and 5 were also examined, but the gain
in goodness of fit was minimal compared to the increase in the number
of free parameters. Therefore, third order polynomials were the best
compromise between complexity and goodness of fit.

The best fitting models were chosen according to their coefficient
of determination (also known as $\mbox{R}^{2}$-value), which is a
statistical measure for the goodness-of-fit \textcolor{green}{\cite{weisstein_correlation_????}}.
It can be simply defined as $R^{2}=1-S_{res}/S_{tot}$, where $S_{res}$
is the residual sum of squares (the residual between the data points
and the fitting function) and $S_{tot}$, which is the total sum of
squares and is proportional to the variance of the data. An $\mbox{R}^{2}$-value
of 1 corresponds to a perfect fit. The analytical expressions for
the best models M2 and M7 are shown in table \ref{tab:Fitting-models},
together with \textcolor{black}{their} $\mbox{R}^{2}$-value. We performed
the whole Fisher analysis (see section \ref{sec:Fisher-Matrix-method}
below), for both of the models and the results on the parameter estimation
are basically the same (less than half of a percent relative difference
in the estimated final errors).

\begin{table}
\centering{}%
\begin{tabular}{|c|c|c|c|}
\hline 
Model Name  & Functional form  & \# of coefficients  & $\mbox{R}^{2}$-value\tabularnewline
\hline 
\hline 
M2  & $f(k)=1+a_{0}+a_{1}\cdot k+c\cdot k\cdot\arctan((k-k_{0})\cdot b)$  & 5  & 0.99996\tabularnewline
\hline 
M7  & $f(k)=1+a_{0}+a_{1}\cdot k+c\cdot k\cdot\frac{b\cdot(k-k_{0})}{\sqrt{1+b^{2}\cdot(k-k_{0})^{2}}}$  & 5  & 0.999989\tabularnewline
\hline 
\end{tabular}\protect\protect\protect\caption{\label{tab:Fitting-models}Fitting models M2 and M7 with their corresponding
number of fitting coefficients and their R$^{2}$-value. Each coefficient
$a_{0}$, $a_{1}$, $c$, \textbf{$b$} and $k_{0}$ is fitted as
a polynomial in the coupling parameter $\beta$ and the redshift $z$
(see Appendix \ref{sec:AppendixA} for further details). The R$^{2}$-value
is a measure of the goodness of fit: a value of 1 corresponds to a
perfect fit, while 0 means that the model does not fit the data.}
 
\end{table}

\subsection{Fitting functions and cosmological parameters\label{sub:Fitt-and-nonlinear-cosmo-pars}}

In order to use the fitting formulae obtained before to forecast cosmological
parameters of the model, we need to assume that the shape of the non-linear
coupled DE signal does not change dramatically if the other cosmological
parameters, apart from $\beta$ and $\sigma_{8}$, are modified by
small amounts. This is justified since in the deeply non-linear regime,
the evolution of perturbations is ruled by mode-mode coupling between
high $k$ wavevectors (non-linear $k$ modes are not independent of
each other anymore), which erases most of the information about initial
conditions and makes the shape of the non-linear power spectrum at
large $k$ practically independent of cosmological background parameters,
such as $n_{s}$, $\Omega_{b}$ and $h$. This has been shown to be
the case when calculating perturbatively non-linear corrections to
the power spectrum, see for example \citep{crocce_memory_2006,pietroni_coarse-grained_2011},
but also analyzing the covariance matrix of non-linear power spectra
using a large suite of N-body simulations as was investigated in \citep{takahashi_non-gaussian_2011}.
Furthermore, the decoupling of virialized structures in the small
scale regime from the background dynamics of the universe, is one
of the cornerstones of the recently developed effective field theory
of large-scale structure \cite{baumann_cosmological_2012-2} and was
also shown to be approximately true using a coarse grained cosmological
perturbation approach \cite{manzotti2014acoarse}. Since $\beta$
and $\sigma_{8}$ are quantities that affect directly the linear perturbations
and the virialization dynamics and we are only looking at a particular
signal at very small scales, they should be the main parameters determining
the shape of the non-linear ``bump''. An investigation of how robust
these fitting formulas are, with respect to a change of cosmological
parameters, would need either more high-resolution N-body simulation
for coupled DE scenarios or the development of a consistent perturbation
theory for modified gravity and scalar-tensor theories that reaches
highly non-linear scales.

\section{Non-linear power spectrum and error estimation\label{sec:Errors-sources}}

\begin{figure}
\centering{}\includegraphics[width=0.45\textwidth]{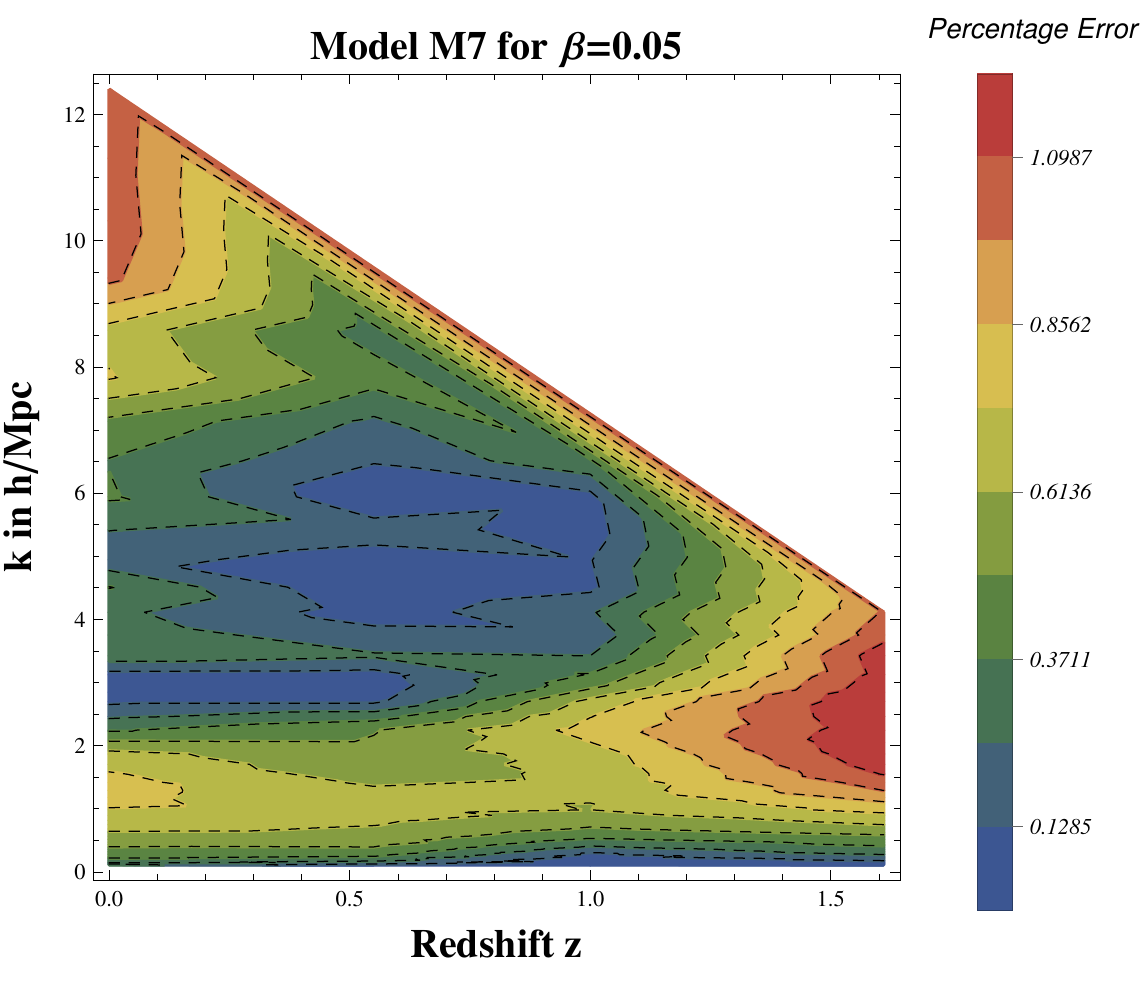}\includegraphics[width=0.45\textwidth]{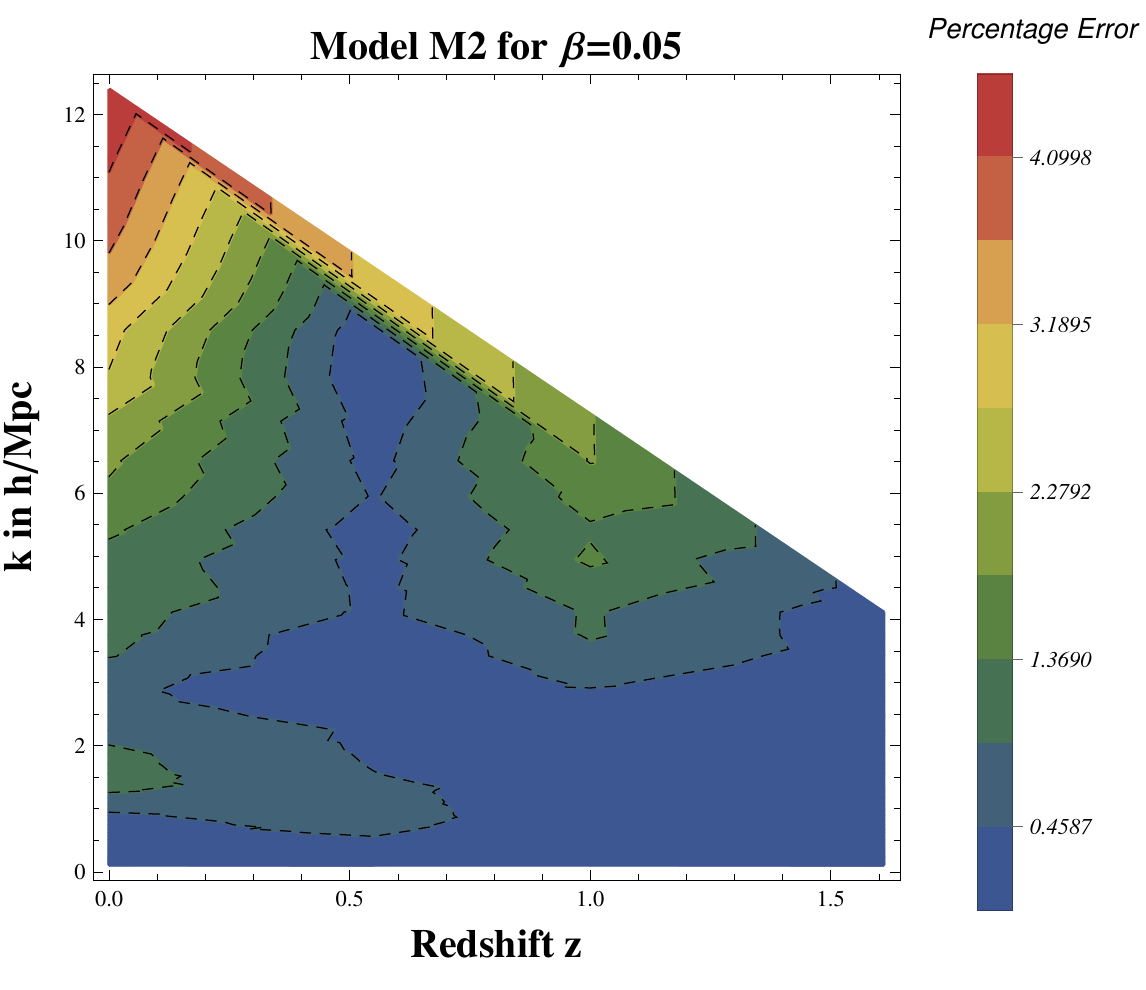}\protect\protect\caption{\label{fig:Error-contour-plot}Error contour plot for the fitting
functions of model M7 (left) and M2 (right) applied compared directly
to the N-body simulations. We show that for the smallest coupling
constant available in the simulations, $\beta=0.05$, the error remains
below 1\% for the scales and redshifts we are interested in.}
\end{figure}

The accuracy of the fitting functions when compared to the original
N-body simulation power spectra is shown in figure \ref{fig:Error-contour-plot},
where the absolute value of the percentage error between fitting function
and the original power spectra is plotted as a function of redshift
and scale. In this case we show that for the model M7 the error remains
well below 1\% for a coupling constant of $\beta=0.05$ and for the
scales and redshifts we are interested in. For higher couplings the
error goes up to a maximum of 1.5\%. For the model M2 the same test
gives similar results, yielding <1\% error in all interesting scales
and redshifts. We can then expect that when applying these fitting
functions on top of $\Lambda$CDM power spectra, the intrinsic error
is less than 1\% and the extra sources of error are all given by the
N-body simulation spectra and by the estimators of the non-linear
$\Lambda$CDM power spectrum.

\begin{figure}
\centering{}\includegraphics[width=0.7\textwidth]{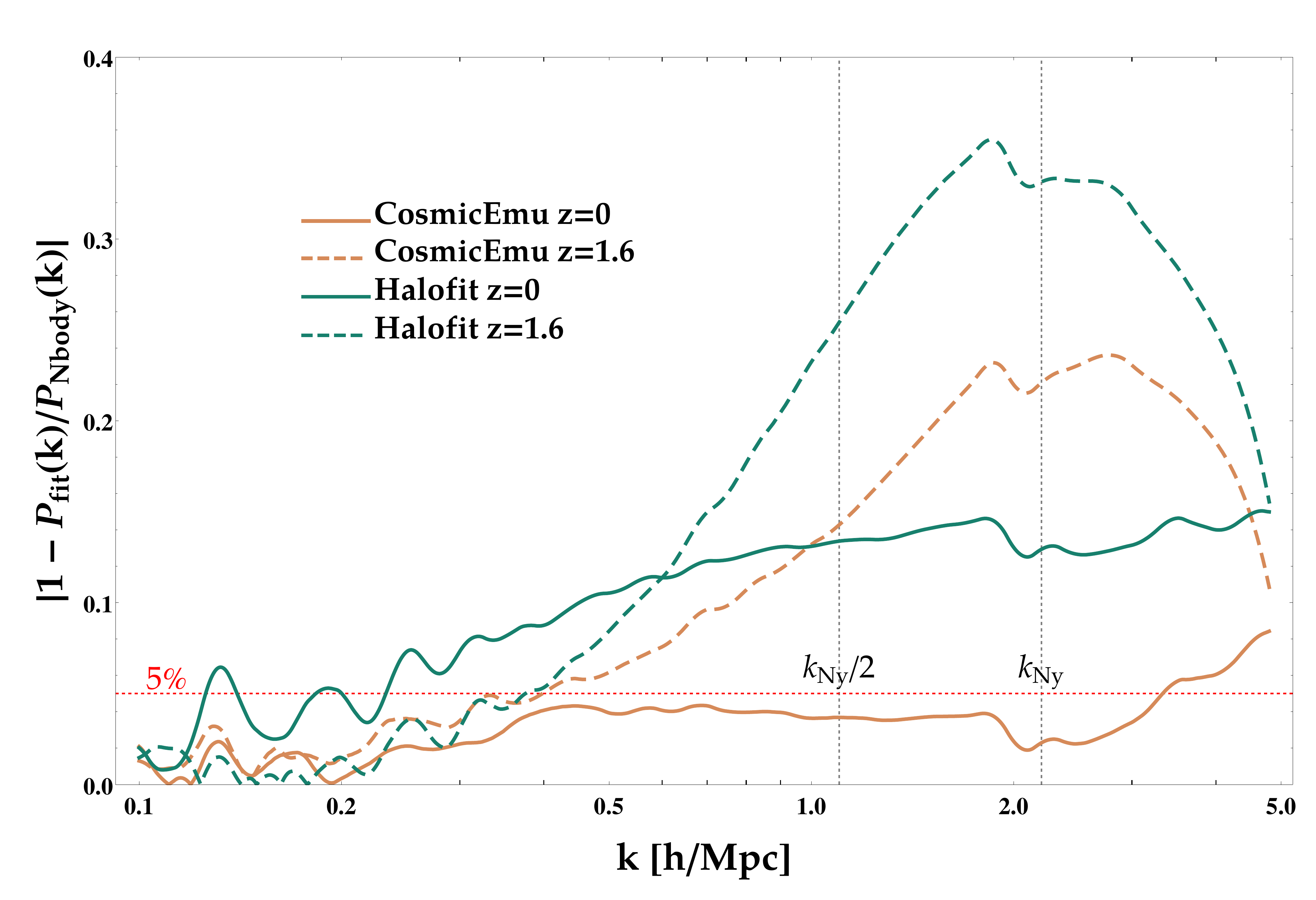}\protect\protect\protect\caption{\label{fig:Error-comp-Halofit-CosmicEmu} Relative error of the fitting
functions Halofit (green) and CosmicEmulator (orange) (FrankenEmu)
with respect to the CoDECS $\protect\lcdm$ N-body power spectra at
two different redshifts (solid lines: $z=0$, dashed lines: $z=1.6$).
While the relative error with respect to the CosmicEmulator remains
below the 5\% limit (horizontal red dashed line) for all interesting
$k$-values, the error compared to Halofit is already bigger than
10\% for $k\apprge0.5$h/Mpc. Both the error of our simulation with
respect to Halofit and to the CosmicEmulator increase as a function
of redshift. The change of trend after the Nyquist frequency (marked
with the vertical grey line) can be attributed to the use of the folding
method for the CoDECS nonlinear power spectra.}
\end{figure}

Since the fitting functions shown in the previous section are useful
when applied on top of a $\Lambda$CDM non-linear power spectrum,
we need to choose an estimator for the non-linear $\Lambda$CDM $P(k)$.
Our tests show that \emph{FrankenEmu }\citep{heitmann_coyote_2014}\emph{,}
an improved version of the original Coyote Cosmic Emulator, performs
better than the revised version of Halofit by \citep{takahashi_revising_2012},
which is included in recent versions of CAMB \citep{lewis_efficient_1999}.
While at $z=0$ both estimators work similarly well with an accuracy
at the 5\% level in the BAO range, Halofit performs much worse with
increasing redshift and increasing $k$, as illustrated in figure
\ref{fig:Error-comp-Halofit-CosmicEmu}. At $z=0$ the Cosmic Emulator
shows a flat error curve for all scales up to the Nyquist frequency,
below the error estimated for Halofit. A comparison between power
spectrum estimators and N-body simulations has been performed also
in \citep{fosalba_mice_2013}, where they found similar results: Halofit
and the Cosmic Emulator perform similarly in the BAO range, but the
errors introduced by Halofit is above the 10\% level at scales of
around $k\approx1\mbox{Mpc}/h$ and $z\apprge1$. Therefore, we use
the \emph{FrankenEmu} as the preferred $\Lambda$CDM non-linear power
spectrum estimator for our forecasting purposes. They claim to be
accurate at the 1-3\% percent level around the scales of interest
and they have performed very careful resolution tests using hundreds
of realizations. 

\begin{figure}
\includegraphics[width=0.7\textwidth]{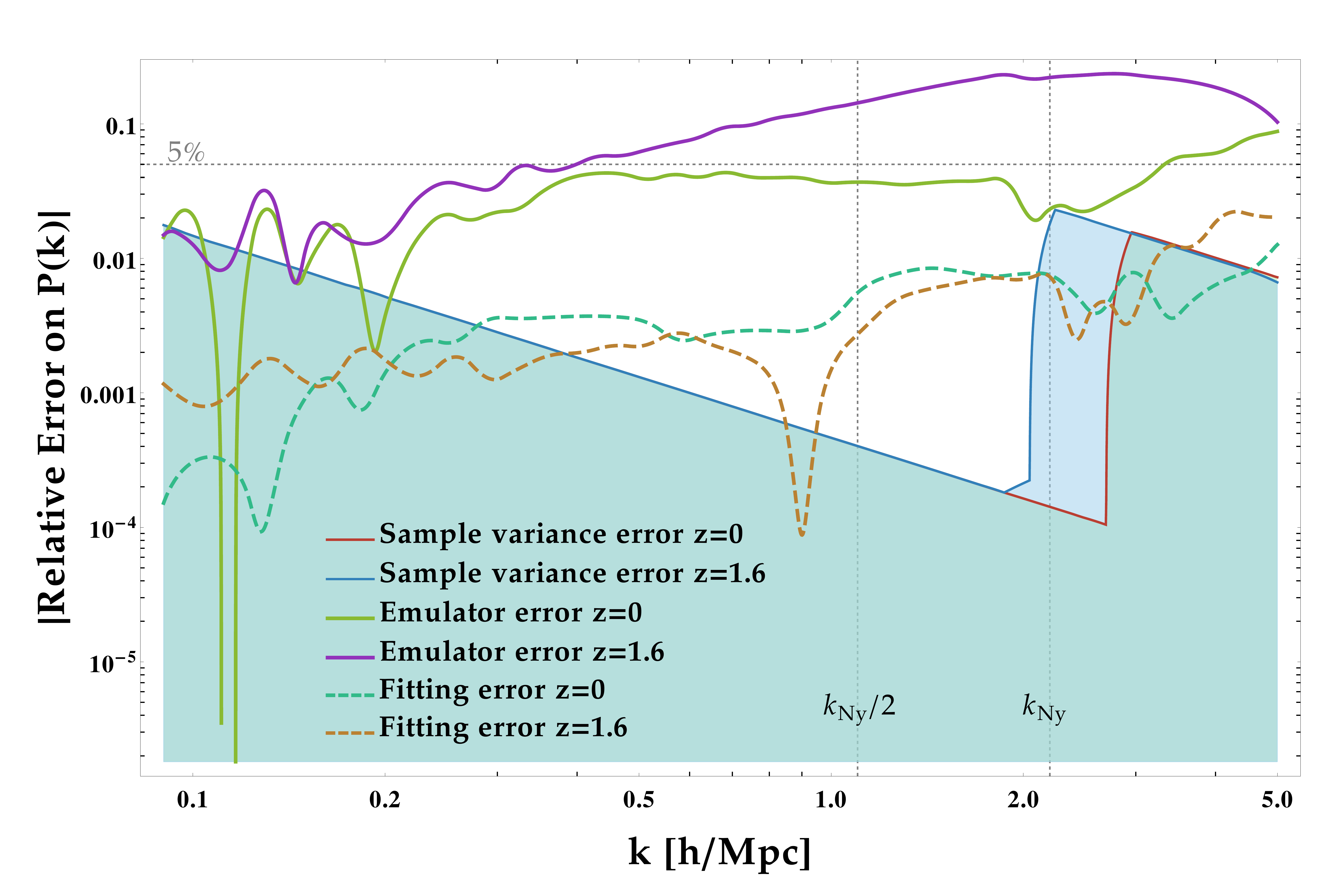}

\protect\caption{\label{fig:Error-sources}Different sources of error affecting the
nonlinear power spectrum. Each source of error is shown at two different
redshifts, $z=0$ and $z=1.6$. The sample variance error (red solid
line and blue solid line, together with their shaded regions) corresponds
to the error induced by the limited number of available $k$-modes
when extracting the power spectrum; it has a sharp increase at the
Nyquist frequency, since there the folded mesh has fewer modes available
than the top level mesh, see section \ref{sec:Folding-method} for
more details. The fitting error (green and orange dashed lines) corresponds
to the intrinsic error of the fitting functions with respect to an
N-body simulation using the same parameters, calculated in section
\ref{sec:Fitting-functions}. The emulator error (green and purple
thick lines) is the error of the Cosmic Emulator compared to a $\protect\lcdm$
N-body simulation from CoDECS (these two lines correspond to the orange
lines of figure \ref{fig:Error-comp-Halofit-CosmicEmu}). The relative
error increases with redshift and scale and reaches more than 15\%
at the Nyquist frequency for the highest redshift. The vertical grey-dashed
lines mark the scales $k_{Ny}/2$ and $k_{Ny}$. }
\end{figure}

Another source of error is the sample variance error of the power
spectrum when extracted from the N-body simulation. This depends on
the number on the number $n_{mod}$ of independent modes available
at each wavevector bin in $k$ and its given by \cite{feldman_power_1994}:
$\sigma_{P}=\sqrt{\frac{2}{n_{mod}}}P(k)$. In figure \ref{fig:Error-sources}
it can be seen as a blue shaded region marked by a red and blue solid
line. For large scales this error is considerable, but there one usually
uses just linear power spectra computed by Boltzmann codes, like CAMB
or CLASS. At $k=0.1h/\mbox{Mpc}$ the binning error is at the percent
level and decreases rapidly to negligible values, but then it increases
again quickly at $k\sim2h/\mbox{Mpc}$ since there the folded mesh
for the power spectrum has again only few modes to sample from, as
explained in section \ref{sec:Folding-method}.

The intrinsic error of the fitting function is shown in figure \ref{fig:Error-sources}
as a dashed green line and it remains well below the 1\% level at
all the scales of interest.

We include all errors discussed here in our Fisher forecast analysis.
It turns out that they do not affect considerably the results for
a survey like Euclid, which will measure such a high number of galaxies,
that the sampling of the clustering signal will be not affected by
small sources of noise in the data.

\section{\label{sec:Fisher-Matrix-method}Fisher Matrix method}

The Fisher matrix formalism (\cite{tegmark_measuring_1998,seo2007improved,seo2005baryonic})
is one of the most popular tools to forecast the outcome of an experiment,
because of its speed and its versatility when the likelihood is approximately
Gaussian (see \cite{sellentin_breaking_2014} for a method on how
to improve this approximation). Here we apply the Fisher matrix formalism
to two different probes, galaxy Clustering (GC) and weak lensing (WL),
which are the ones that a future satellite like Euclid \cite{amendola_cosmology_2012-short}
will carry on board.

\subsection{Galaxy Clustering}

The Fisher matrix for the galaxy power spectrum has the following
form \citep{seo2007improved, amendola2012testing}:

\begin{equation}
F_{ij}=\frac{V_{survey}}{8\pi^{2}}\int_{-1}^{+1}\mbox{d}\mu\int_{k_{min}}^{k_{max}}\mbox{d}k\, k^{2}\frac{\partial\ln P_{obs}(k,\mu,z)}{\partial\theta_{i}}\frac{\partial\ln P_{obs}(k,\mu,z)}{\partial\theta_{j}}\left[\frac{n(z)P_{obs}(k,\mu,z)}{n(z)P_{obs}(k,\mu,z)+1}\right]^{2}\,\,.\label{eq:fisher-matrix-gc}
\end{equation}

Here $V_{survey}$ is the volume covered by the survey and contained
in a redshift slice $\Delta z$, $n(z)$ is the galaxy number density
as a function of redshift, $P_{obs}(k,\mu,z)$ is the observed power
spectrum as a function of the redshift $z$, the wavenumber $k$ and
of $\mu\equiv\cos\alpha$, where $\alpha$ is the angle between the
line of sight and the 3D-wavevector $\vec{k}$.

The derivatives in eq. (\ref{eq:fisher-matrix-gc}) are taken with
respect to a vector of cosmological parameters, $\theta_{i}=\{\beta^{2},\, h,\,\log\mathcal{A}_{s},\, n_{s},\,\omega_{c},\,\omega_{b},\, b(z)\}$,
where $b(z)$ is the galaxy bias. When we perform the numerical derivatives
of $P_{obs}(k,\mu,z)$, we have to take into account that not only
the power spectrum and the background functions change, when the cosmological
parameters are modified, but also $\mu$ and $k$ are changed by a
geometrical factor depending on $H(z)$ and $D_{A}(z)$, due to the
Alcock-Paczinsky effect \citep{alcock1979anevolution}.

The observed power spectrum contains all the cosmological information
about the background and the matter perturbations as well as corrections
due to redshift-space distortions and observational uncertainties.
In this work we write it in the following form:

\begin{equation}
P_{obs}(k,\mu,z)=\frac{D_{A,f}^{2}(z)H(z)}{D_{A}^{2}(z)H_{f}(z)}b^{2}(z)(1+\beta_{d}(z)\mu^{2})^{2}e^{-k^{2}\mu^{2}(\sigma_{r}^{2}+\sigma_{v}^{2})}P(k,z)\label{eq:observed-Pk}
\end{equation}

In the above formula, the subscript $f$ denotes the fiducial value
of each quantity, $P(k,z)$ is the matter power spectrum, $D_{A}(z)$
is the angular diameter distance, $H(z)$ the Hubble function and
$\beta_{d}(z)$ is the redshift space distortion factor, which in
linear theory is given by $\beta_{d}(z)=f(z)/b(z)$, with $f(z)\equiv d\ln G/d\ln a$
representing the linear growth rate of matter perturbations. The exponential
factor represents the damping of the observed power spectrum, due
to two different effects: $\sigma_{z}$ an error induced by spectroscopic
redshift measurement errors, which translates into a damping scale
$\sigma_{r}=\sigma_{z}/H(z)$ and $\sigma_{v}$ which is the dispersion
of pairwise peculiar velocities which are present at non-linear scales
and also introduce a damping scale in the mapping between real and
redshift space.

\subsubsection{Including sources of error in the Fisher formalism\label{sub:Including-sources-of}}

In section \ref{sec:Errors-sources} we discussed several sources
of error affecting the non-linear power spectrum, the intrinsic errors
of the coupled DE fitting functions, mode-binning errors in the N-body
power spectrum and the estimation and interpolation error of the $\Lambda\mbox{CDM}$
non-linear power spectrum obtained from the CosmicEmulator. We will
take these sources of error into account in our Fisher forecast analysis
by including them as extra noise affecting the observed power spectrum.
The term in square brackets in eq. (\ref{eq:fisher-matrix-gc}) corresponds
to the inverse of the covariance $C=P(k,z)+n(z)^{-1}$. The ``noise
term'' $n(z)^{-1}$ is the number density of sampling points for
the matter power spectrum (galaxies in a survey), which gives us an
estimate of the signal-to-noise ratio we can expect from the forecast:
for a higher number density, the power spectrum is better sampled
and more information can be extracted from it. In order to take into
account the theoretical and numerical errors on $P(k,z)$, we decrease
$n(z)$ by a factor that contains the relative errors on $P(k,z)$.
In eq. (\ref{eq:fisher-matrix-gc}), instead of $n(z)$, we then have
an ``effective'' number density:

\begin{equation}
n_{eff}(k,z)=n(z)/(1+n(z)\tilde{\sigma}_{p}(k,z))
\end{equation}
The term $\tilde{\sigma}_{p}(k,z)$ is a scale- and redshift-dependent
term which is the square root of the sum of the relative errors squared.
We take into account all error sources which affect $P(k,z)$ due
to different reasons, as explained in section \ref{sec:Errors-sources}.
One of them is the difference between our power spectrum estimator
and the N-body simulation: $\sigma_{p}(k,z)=(P_{numerical}(k,z)-P_{true}(k,z))/P_{true}(k,z)$.
If $\tilde{\sigma}_{p}(k,z)=0$ or it is negligible, the effective
number density will be the observed one $n_{eff}(z)=n(z)$; otherwise,
$n_{eff}(z)<n(z)$, the effective number of sampling points being
reduced, together with the amount of information one can extract from
the power spectrum. As long as $n(z)P(k,z)\gg1$ for all $z$ and
$k$, the sampling will be always good enough to extract cosmological
information from the power spectrum even in the presence of noise.
For the specifications used in this work (see table \ref{tab:GC-specifications}
below), $n(z)P(k,z)$ is larger than 1 in all scales of interest and
therefore the theoretical and numerical error on $P(k,z)$ does not
have such a considerable effect on the parameter estimation, as one
would expect naively. We test the inclusion of the effective number
density $n_{eff}(z)$ on the Fisher forecast analysis and we find
that the relative 1-$\sigma$ marginalised errors on each cosmological
parameter are between 8\% and 15\% higher when using the estimated
uncertainties on the power spectrum.

\subsubsection{Systematic bias on cosmological parameters}

In this section we will quantify the effect of the systematic errors
due to the uncertainties on the non-linear power spectrum. We will
show how big this systematic bias would be, if we used for our forecasts
a power spectrum which is not the ``correct'' one, for example one
obtained from a single N-body realization. For the following discussion
on systematic biases, we will use the expressions derived in the Appendix
B of \cite{taylor_probing_2007}.

The linear bias on a cosmological parameter $\delta\theta_{i}$ due
to the bias $\delta\psi_{i}$ in a parameter of the model which we
assume fixed and cannot be measured is given by: 

\begin{equation}
\delta\theta_{i}=-\left[F^{\theta\theta}\right]_{ik}^{-1}F_{kj}^{\theta\psi}\delta\psi_{j}\label{eq:syst.bias}
\end{equation}

In our case we will have only one systematic parameter $\psi$, which
controls the difference between the ``true'' power spectrum $P_{true}$
and and our simulated power spectrum $P_{num}$:

\[
P_{\psi}=\psi P_{num}+(1-\psi)P_{true}
\]
$\psi$ can vary continously so that for $\psi=1$ we recover $P_{num}$,
while for $\psi=0$ we obtain $P_{true}$. We can define 
the relative difference between $P_{true}$ and $P_{num}$ as:
\begin{equation}
\sigma_{p}(k,z)\equiv\frac{P_{num}(k,z)-P_{true}(k,z)}{P_{true}(k,z)}
\end{equation}

The $F^{\theta\theta}$ in eqn.\ref{eq:syst.bias} above is simply
the usual Fisher matrix: 
\[
F^{\theta\theta}=\frac{1}{2}\mbox{tr}\left(C^{-1}\partial_{i}^{\theta}CC^{-1}\partial_{j}^{\theta}C\right)
\]
while the pseudo-Fisher matrix between measured and assumed parameters
$F^{\theta\psi}$ is:

\begin{equation}
F_{ij}^{\theta\psi}=\frac{1}{2}\mbox{tr}\left(C^{-1}\partial_{i}^{\theta}CC^{-1}\partial_{j}^{\psi}C\right)
\end{equation}
which for one systematic parameter only, is just a column vector.

In the case of galaxy Clustering we will compute $F^{\theta\psi}$
in the following way, using the fact that for $\psi=1$, $C=P_{num}(k,z)+n^{-1}(z)$
and $P_{\psi}|_{\psi=1}=P_{num}$:

\begin{equation}
F_{i}^{\theta\psi}\propto\int\mbox{d}k\, k^{2}\left(\frac{n_{eff}(z)P_{num}(k,z)}{n_{eff}(z)P_{num}(k,z)+1}\right)^{2}\left(\frac{1}{P_{num}(k,z)}\right)^{2}\left.\frac{\partial P_{\psi}}{\partial\psi}\right|_{\psi=1}\frac{\partial P_{num}}{\partial\theta_{i}}\label{eq:pseudo-Fisher}
\end{equation}

in this step we have assumed that we have no systematic parameters
affecting the galaxy number density $n(z)$ and therefore, its derivative
w.r.t $\psi$ vanishes. Also, for notational simplicity we left out the integral over $\mu$ and the
form of the observed power spectrum.

We have then $\partial_{\psi}P_{\psi}=-P_{true}+P_{num}=P_{true}\sigma_{p}(k,z)$
and we just need to perform eqn. \ref{eq:pseudo-Fisher} with eqn.
\ref{eq:syst.bias} in order to obtain the systematic biases on the
cosmological parameters.

In the following table \ref{tab:systbias} we present the results
on the systematic bias, for a standard $\lcdm$ forecast, for different
maximum $k$ coverages, up to a maximum $k$ of $1.1$h/Mpc. We will
regard as a ``true'' power spectrum $P_{true}$, the one obtained
by the Cosmic (Franken) Emulator \cite{heitmann_coyote_2014}, since
they have performed a careful analysis of resolution effects using
a large set of simulations and claim to be accurate for $k<1.0\,\mbox{h/Mpc}$
at the 1\% percent level compared to state of the art N-body simulations.
On the other hand, the numerical ``biased'' power spectrum $P_{num}$,
is the one obtained from the CoDECS $\lcdm$ run, which consists on
only one realization. We have left out the CDE coupling parameter
$\beta$, since in that case we do not have any other prediction in
the non-linear regime to compare with. 

\begin{table}[H]
\centering{}%
\begin{tabular}{|c|c|c|c|c|c|}
\hline 
\textbf{Parameter}  & $h$  & $\ln\mathcal{A}_{s}$  & $n_{s}$  & $\omega_{b}$  & $\omega_{c}$\tabularnewline
\hline 
fiducial  & 0.7036  & 2.42  & 0.966  & 0.04503  & 0.2256\tabularnewline
\hline 
\textbf{$\mathbf{k_{max}=k_{Ny}/2=1.1\mbox{\textbf{h/Mpc}}}$}  &  &  &  &  & \tabularnewline
syst. bias with $\delta\psi=1$ & -0.0016 & -0.15 & 0.061 & 0.0028 & -0.0031\tabularnewline
\hline 
\textbf{$\mathbf{k_{max}=0.35\mbox{\textbf{h/Mpc}}}$}  &  &  &  &  & \tabularnewline
\hline 
syst. bias with $\delta\psi=1$ & -0.0094 & -0.11 & 0.045 & 0.0021 & -0.0039\tabularnewline
\hline 
\textbf{$\mathbf{k_{max}=0.15\mbox{\textbf{h/Mpc}}}$}  &  &  &  &  & \tabularnewline
\hline 
syst. bias with $\delta\psi=1$ & -0.0032 & -0.04 & 0.018 & 0.00026 & -0.0024\tabularnewline
\hline 
\end{tabular}\protect\caption{\label{tab:systbias} Systematic bias on the $\protect\lcdm$ cosmological
parameters evaluated at our fiducial model. The systematic bias is
higher when probing smaller scales, where uncertainties in the non-linear
power spectrum are higher. This highlights the importance of modelling
correctly the non-linear power spectrum in order to analyze data.
Using the wrong non-linear power spectrum produces statistical errors
which are larger or of the same order as the statistical results. }
\end{table}

The results of table \ref{tab:systbias} show how important it is
to model accurately the non-linear power spectrum in order to make
forecasts and to analyze the upcoming data of large scale structure
surveys like Euclid \cite{amendola_cosmology_2012-short}. The systematic
errors on the cosmological parameters can be bigger than the statistical
errors (compare to table \ref{tab:1-sigmaerrors-nl-GC-1} and figure
\ref{fig:kmax-change-lin-nonlin} in the results section below). This
is the case if, as in our example scenario, we would use a non-linear
power spectrum that is inaccurate by about 10-15\% in the non-linear
regime at higher redshifts (which was shown in figure \ref{fig:Error-comp-Halofit-CosmicEmu}).
Therefore, it is well justified to choose for our Fisher forecasts
the Cosmic Emulator as the ``true'' $\lcdm$ non-linear power spectrum
estimator, since this is the most accurate predictor up to date. It
would still be interesting to know how robust is the signal of the
coupling parameter $\beta$ with respect to changes in the other parameters
or in the $\lcdm$ non-linear prediction, but as long as we do not
have better and faster semianalytic methods applicable to general
models of dark energy, the estimation of systematic biases of extra
parameters is an impossible task.

\subsubsection{Choice of the $k_{max}$ integration limit in the Fisher formalism}

There are at least three ways of setting the maximum $k$ mode used
in the Fisher forecast integration (eqn. \ref{eq:fisher-matrix-gc}).
One common choice is to set a hard cut in $k$ at all redshifts, and
depending if one wants to include or not non-linear effects, this
cut can be taken at linear scales, smaller than $k=0.1h/\mbox{Mpc}$,
or at non-linear scales $k>0.1h/\mbox{Mpc}$. In the latter case if
one is using a power spectrum calculated in linear theory, one needs
to to use some Lagrangian damping correction, as introduced originally
in \cite{seo2007improved,seo_nonlinear_2008}, in order to take into
account broadening effects on the BAO peak induced by the non-linear
evolution of densities and velocities. Another option to cut off the
power spectrum is to demand that the variance $\sigma_{8}(z;k)$ stays
below a specified value at each redshift, therefore implicitly changing
the cutoff in $k$ as a function of $z$. An usual choice for this
is $\sigma_{8}(z;k_{cut})=0.25$, as was done previously in \cite{amendola_testing_2012}.
Since we are assuming to have a knowledge of the non-linear power
spectrum up to the Nyquist frequency, we implement for our forecasts
a hard-cut method, at $k_{Ny}$ and at $k_{Ny}/2$, without the need
of any Lagragian damping correction. However, to be conservative,
we cite as our main results the ones in which the cut is performed
at $k_{Ny}/2$ in order to eliminate any possible unknown contribution
from the numerical high-frequency noise entering the estimation of
$P(k)$ at the Nyquist frequency (see e.g. \cite{colombi_accurate_2009,Fosalba2013}
for similar prescriptions on where to cut the non-linear power spectrum)\textcolor{black}{.}

\subsection{Weak Lensing \label{sub:Weak-Lensing}}

For the weak lensing power spectrum the Fisher matrix is just a sum
over all possible correlations at different redshift bins \citep{tegmark_measuring_1998},
namely :

\begin{equation}
F_{\alpha\beta}=f_{sky}\sum_{\ell}^{\ell_{max}}\sum_{i,j,k,m}\frac{(2\ell+1)\Delta\ell}{2}\frac{\partial P_{ij}(\ell)}{\partial\theta_{\alpha}}C_{jk}^{-1}\frac{\partial P_{km}(\ell)}{\partial\theta_{\beta}}C_{mi}^{-1}\label{eq:FisherSum-WL}
\end{equation}
With the power spectrum of the shear field given by:

\begin{equation}
P_{ij}(\ell)=\frac{9}{4}\int_{0}^{\infty}\mbox{d}z\:\frac{W_{i}(z)W_{j}(z)H^{3}(z)\Omega_{m}^{2}(z)}{(1+z)^{4}}P_{m}(\ell/r(z))
\end{equation}
where $P_{m}$ is the matter power spectrum discussed above, the indices
$i,\: j$ stand for each of the $\mathcal{N}$ redshift bins and the
window functions are given by:

\begin{equation}
W(z)=\int_{z}^{\infty}\mbox{d}\tilde{z}\left(1-\frac{r(z)}{r(\tilde{z})}\right)n(\tilde{z})
\end{equation}
where the normalized galaxy distribution function is:

\begin{equation}
n(z)=z^{2}\exp\left(-(z/z_{0})^{3/2}\right)\label{eq:ngal dist}
\end{equation}
Here the median redshift and $z_{0}$ are related by $z_{med}\approx1.412z_{0}$.
The corresponding covariance matrix has the following form:

\begin{equation}
C_{ij}(\ell)=P_{ij}(\ell)+\delta_{ij}\gamma_{int}^{2}n_{i}^{-1}+K_{ij}(\ell)
\end{equation}
where $\gamma_{int}$ is the intrinsic galaxy ellipticity and the
shot noise term contains

\begin{equation}
n_{i}=3600\left(\frac{180}{\pi}\right)^{2}n_{\theta}/\mathcal{N}
\end{equation}
with $n_{\theta}$ the total number of galaxies per arcmin\texttwosuperior{}
assuming that the redshift bins have been chosen such that each contain
the same amount of galaxies (equi-populated bins). The matrix $K_{ij}(\ell)$
is a diagonal ``cutoff'' matrix whose entries increase to very high
values at the scales where the power spectrum $P(k)$ has to be cut
to avoid the inclusion of numerical noise errors. Since $K_{ij}(\ell)$
depends on the multipole $\ell$, which itself depends on $k$ and
$z$ through: $\ell=k\, r(z)$, each of the entries $K_{ii}$ increase
with the multipole in a different way. For our purposes we chose a
very sharp cutoff, such that multipoles containing wavenumbers $k$
beyond the Nyquist frequency (or half of the Nyquist frequency, for
our reference case) at the center of the redshift bin $i$, do not
contribute to the Fisher matrix at all after the corresponding $\ell_{cut}$
(specified in table \ref{tab:WL-zbins-specs} below).

\subsection{Adding non-linear corrections to the power spectrum\label{sub:Adding-non-linear-corrections}}

We are interested in a Fisher forecast that includes information from
non-linearities. In this case we cannot separate the power spectrum,
into a power spectrum at redshift zero multiplied by the square of
the normalized growth factor, but we need to evaluate directly the
non-linear power spectra $P(k,z)$ at different redshifts. A full
non-linear correction for the redshift space distortions would be
also desirable, but since the modelling of that effect is yet to be
understood in general cases, we use as a first approximation an exponential
damping of the form $\exp(-k^{2}\mu^{2}\sigma_{v}^{2})$, where $\sigma_{v}$
is the pairwise peculiar velocity of galaxies induced by non-linearities
in the matter and velocity power spectrum. This is the first term
of a set of corrections that can be applied to the Kaiser formula
\citep{kaiser_clustering_1987} for the clustering in redshift space
(see e.g. \cite{de_la_torre_modelling_2012}, \cite{scoccimarro_redshift-space_2004},
\cite{taruya_baryon_2010},\cite{wang_toward_2012}). We use a value
of $\sigma_{v}=300\mbox{km}/\mbox{s}$ which is an approximate and
conservative value based on the estimations by \citep{de_la_torre_modelling_2012},
in which the authors test a variety of redshift-space distortion models.

As already mentioned, the damping term, introduced by \citep{seo2007improved},
which should correct the linear $P(k)$, especially the position of
the BAO peaks, for non-linearities, is not included here, since we
assume that we have a complete model of the non-linear power spectrum
and therefore all possible corrections are already included in our
fitting functions and power spectrum emulators.

In order to implement our model of coupled Dark Energy, we use for
$H(z)$ and $D(z)$ tables precomputed using a modified version of
CAMB, that includes the exponential and inverse-power law potentials
for coupled dark energy \cite{amendola_testing_2012,pettorino_testing_2013}.
We calculate these tables for each of the parameters $\theta_{i}\pm\epsilon$,
where $\epsilon=0.03.$ \textcolor{black}{We do the same for the linear
perturbation quantity $G(z)$, whose logarithmic derivative with respect
to $\ln(a)$, known as the growth rate $f(z)$ enters the redshift
space distortion term in \ref{eq:observed-Pk}. The background quantities
}$H(z)$ and $D(z)$ are important in the Fisher forecast for the
Alcock-Paczynski geometrical term, which we take into account in the
observed power spectrum.

The full non-linear power spectrum we use in our method is then obtained
as follows. 
\begin{itemize}
\item At linear scales $k\apprle0.1h/\mbox{Mpc}$ the linear power spectrum
is obtained from our modified version of CAMB which includes the effect
of the DM-DE coupling. 
\item At non-linear scales, $k\apprge0.1h/\mbox{Mpc}$, we use a combination
of the power spectrum calculated with the cosmic emulator \emph{FrankenEmu
}and our fitting formulas for CDE that account for the non-linear
dynamics in presence of a fifth force. 
\item The matching at $k\approx0.1h/\mbox{Mpc}$ is performed using different
interpolation methods, either averaging on both sides and smoothing
out or allowing for a small discontinuity. The matching point is left
out of the Fisher integral and we check for this effect, showing that
the different methods have a negligible effect (less than 2 percent)
on the final absolute errors on the parameters. 
\item In order to be conservative in terms of numerical errors and noise,
we cut the power spectra at half of the Nyquist frequency (we also
test and compare with a cutoff at $k_{Ny}$) and we include all sources
of errors specified in section \ref{sub:Including-sources-of} , as
effective noise terms into our Fisher estimation. 
\end{itemize}

\section{\label{sec:GC-WL-Results}Results}

We now present results for galaxy clustering and weak lensing, using
the specifications for a Euclid-like survey as described in tables
\ref{tab:GC-specifications} and \ref{tab:WL-specifications-1} below.
We use the Fisher formalism described in section \ref{sec:Fisher-Matrix-method}
to forecast the errors in the cosmological parameters, using information
from the non-linear power spectrum for coupled Dark Energy models,
as obtained with the procedure described in section \ref{sec:Folding-method}
together with the fitting functions obtained in section \ref{sec:Fitting-functions}.
To make our estimation more realistic, we also take into account all
sources of error and systematics for the power spectrum which were
discussed in section \ref{sec:Errors-sources}. As a way of testing
our improvement on the parameter estimation, we also perform two extra
Fisher forecasts, first using only linear power spectra for the CDE
model and then correcting these linear $P(k)$ with the latest version
of Halofit \cite{takahashi_revising_2012}, which was designed for
$\lcdm$ only.

The fiducial parameters are $\omega_{c}=0.1117$, $\omega_{b}=0.0223$,
$n_{s}=0.966$, $\log\mathcal{A}_{s}=-19.8395$, $h=0.7036$, $\beta^{2}=0.0025$,
which are consistent with WMAP7 results (see table \ref{tab:ParametersWMAP7}).
The fiducial values for the galaxy bias $b(z)$ used for the GC probe
are taken from the Euclid specifications (see \cite{amendola_cosmology_2012-short,laureijs_euclid_2011,orsi2010probing})
For our final results, we convert these parameters into the set $p_{i}=\{\beta^{2},\, h,\,10^{9}\mathcal{A}_{s},\, n_{s},\,\Omega_{c},\,\Omega_{b}\}$
marginalizing over the bias $b(z)$ (for the GC case) and using a
Jacobian transformation to convert into the new set of parameters,
which is allowed by the Fisher matrix formalism. We choose to forecast
the error on the square of the coupling parameter $\beta^{2}$, because
this is the quantity entering the modified gravitational Newton constant
$G_{eff}$ in the limit of linear perturbations (cf. eq. \ref{linear_cdm}),
therefore giving the strength of the ``fifth force''. The corresponding
fiducial value for the coupling, $\beta=0.05$, is still compatible
with recent limits set by analyzing the data from the Planck Satellite
(see \cite{pettorino_testing_2013,planckcollaboration_planck2015_2015}).

\subsection{Galaxy clustering}

\begin{table}
\centering{}%
\begin{tabular}{|c|c|c|}
\hline 
\textbf{Parameter}  & \textbf{Value}  & \textbf{Description}\tabularnewline
\hline 
$A_{survey}$  & 15000 $\mbox{deg}^{2}$  & Survey area in the sky\tabularnewline
$\sigma_{z}$  & 0.001  & Photometric redshift error\tabularnewline
$\sigma_{v}$  & 300 km/s  & Fiducial pairwise velocity dispersion\tabularnewline
$\Delta z$  & 0.2  & Redshift bin width\tabularnewline
$\{z_{min},\ z_{max}\}$  & \{0.6, 2.0\}  & Min. and max. limits for redshift bins \tabularnewline
\hline 
\end{tabular}\protect\protect\caption{\label{tab:GC-specifications} Specifications for the Fisher Matrix
of an Euclid-like galaxy clustering survey.}
\end{table}

Table \ref{tab:GC-specifications} shows the specifications of a Euclid-like
survey, which are used in our Fisher forecast. While Euclid specifications
use 14 redshift bins of a width $\Delta z=0.1$ (see \cite{amendola_cosmology_2012-short},
table 3 in that work), we use only 6 bins of a width of 0.2. We check
in the case of $\lcdm$, that this re-binning ( done using the specified
number of galaxies and the corresponding comoving volumes in our cosmology)
has a very small effect (of a few percent) on the estimation of the
1-$\sigma$ errors on the parameters.

In table \ref{tab:1-sigmaerrors-nl-GC-1} we show the fully marginalized
1-$\sigma$ errors on the final cosmological parameters $p_{i}=\{\beta^{2},\, h,\,10^{9}\mathcal{A}_{s},\, n_{s},\,\Omega_{c},\,\Omega_{b}\}$,
performing the non-linear power spectrum cut-off at two different
scales: $k_{max}=k_{Ny}/2$ and $k_{max}=k_{Ny}$. As explained above,
we take as a reference result the one corresponding to a cutoff at
$k_{Ny}/2$. The gain in constraining the $\beta^{2}$ parameter when
going from $k_{Ny}/2$ to $k_{Ny}$ is of two percent points in the
relative errors, while for the other cosmological parameters the improvement
is negligible.

\begin{table}
\centering{}%
\begin{tabular}{|c|c|c|c|c|c|c|}
\hline 
\textbf{Parameter}  & \textbf{$\text{\ensuremath{\beta}}^{2}$}  & $h$  & $10^{9}\mathcal{A}_{s}$  & $n_{s}$  & $\Omega_{b}$  & $\Omega_{c}$\tabularnewline
\hline 
fiducial  & 0.0025  & 0.7036  & 2.42  & 0.966  & 0.04503  & 0.2256\tabularnewline
\hline 
\textbf{$\mathbf{k_{max}=k_{Ny}/2}$}  &  &  &  &  &  & \tabularnewline
\hline 
abs. error  & 0.000346  & 0.00160  & 0.01855  & 0.00267  & 0.00084  & 0.00088\tabularnewline
relative error & 14\%  & 0.23\%  & 0.77\%  & 0.28\%  & 1.9\%  & 0.39\%\tabularnewline
\hline 
\textbf{$\mathbf{k_{max}=k_{Ny}}$}  &  &  &  &  &  & \tabularnewline
\hline 
abs. error  & 0.000305  & 0.00151  & 0.01808  & 0.00249  & 0.00081  & 0.00085\tabularnewline
relative error & 12\%  & 0.22\%  & 0.75\%  & 0.26\%  & 1.8\%  & 0.38\%\tabularnewline
\hline 
\end{tabular}\protect\protect\caption{\label{tab:1-sigmaerrors-nl-GC-1} 1-$\sigma$ fully marginalized
errors on the cosmological parameters of the model for a galaxy clustering
Fisher forecast. Two cases are presented, a hard cutoff of the power
spectrum at $k_{max}=k_{Ny}/2$, and a hard cut at $k_{Ny}$.}
\end{table}

\subsubsection{Variation of the $k_{max}$ integration limit}

\begin{figure}[h]
\centering{}\includegraphics[width=0.7\textwidth]{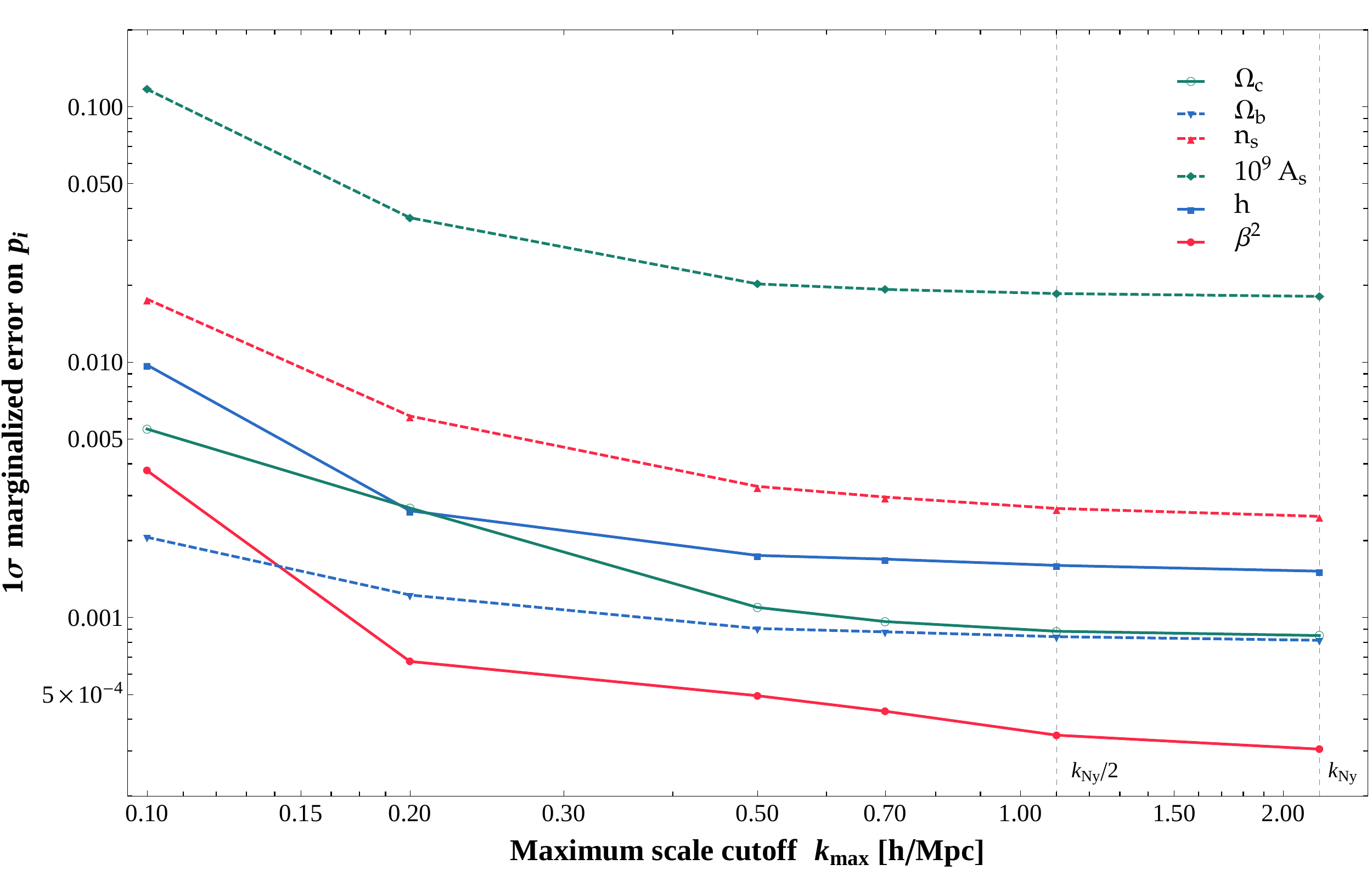}\protect\protect\caption{\label{fig:kmax-change-lin-nonlin} Change of the 1-$\sigma$ fully
marginalized error on the set of cosmological parameters $p_{i}$
for a galaxy clustering Fisher forecast, as a function of the maximum
mode $k_{max}$ used as a cutoff in the Fisher matrix integration.
When increasing the maximum $k_{max}$ the errors on the parameters
get steadily smaller, especially when going from linear ($k\approx0.1h/\mbox{Mpc}$)
to mildly non-linear ($k\approx0.3h/\mbox{Mpc}$) scales. The vertical
dashed grey lines, mark the half and the full Nyquist frequencies.}
\end{figure}

We now test the gain in information obtained by including progressively
more non-linear wavemodes $k$ into the Fisher integration. We perform
the same Fisher forecast, each time increasing the maximum mode $k_{max}$
at which the integration is cut off. In figure \ref{fig:kmax-change-lin-nonlin}
we show how the 1-$\sigma$ fully marginalized error on the cosmological
parameters $p_{i}$ changes with an increase of $k_{max}$. The error
decreases steadily with an increase of $k_{max}$, where the biggest
gain is achieved when going from linear ($k\approx0.1h/\mbox{Mpc}$)
to mildly non-linear ($k\approx0.3h/\mbox{Mpc}$) scales. For parameters
like $h$ and $\Omega_{b}$, an approximate plateau is reached already
before $k_{max}\approx1.0$, while for $\beta^{2}$ there is still
a considerable gain in parameter constraints when going into smaller
scales, even beyond $k_{Ny}/2$ (consistent with table \ref{tab:1-sigmaerrors-nl-GC-1}
above). This happens, qualitatively, because at small scales we have
a well-defined characteristic signal coming from non-linear interactions
including the fifth force which itself involves the $\beta$ coupling,
while the information on the backgorund cosmological parameters gets
washed out (c.f. section \ref{sub:Fitt-and-nonlinear-cosmo-pars}
above).

\begin{figure}
\centering{}\includegraphics[height=0.35\textheight]{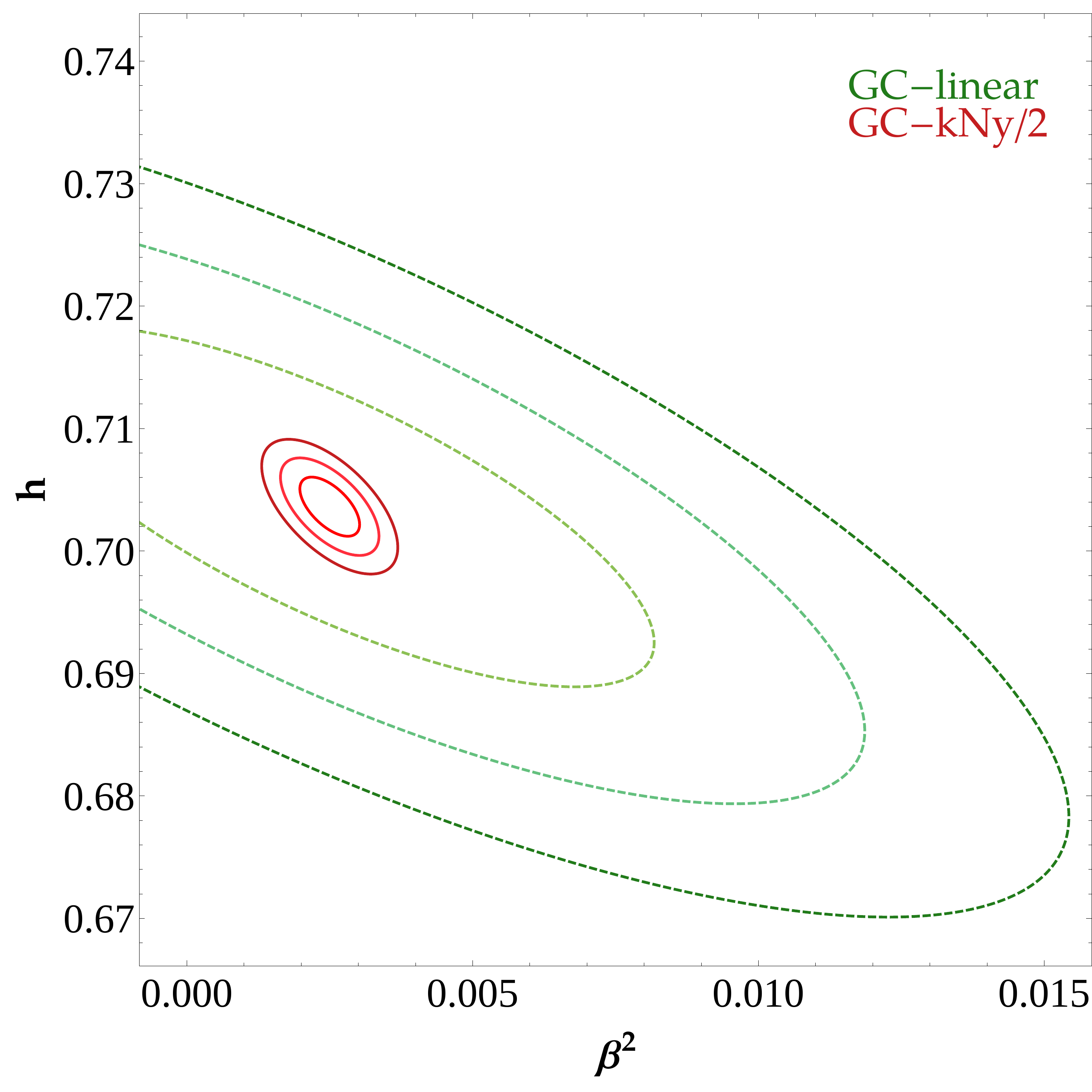}\includegraphics[height=0.35\textheight]{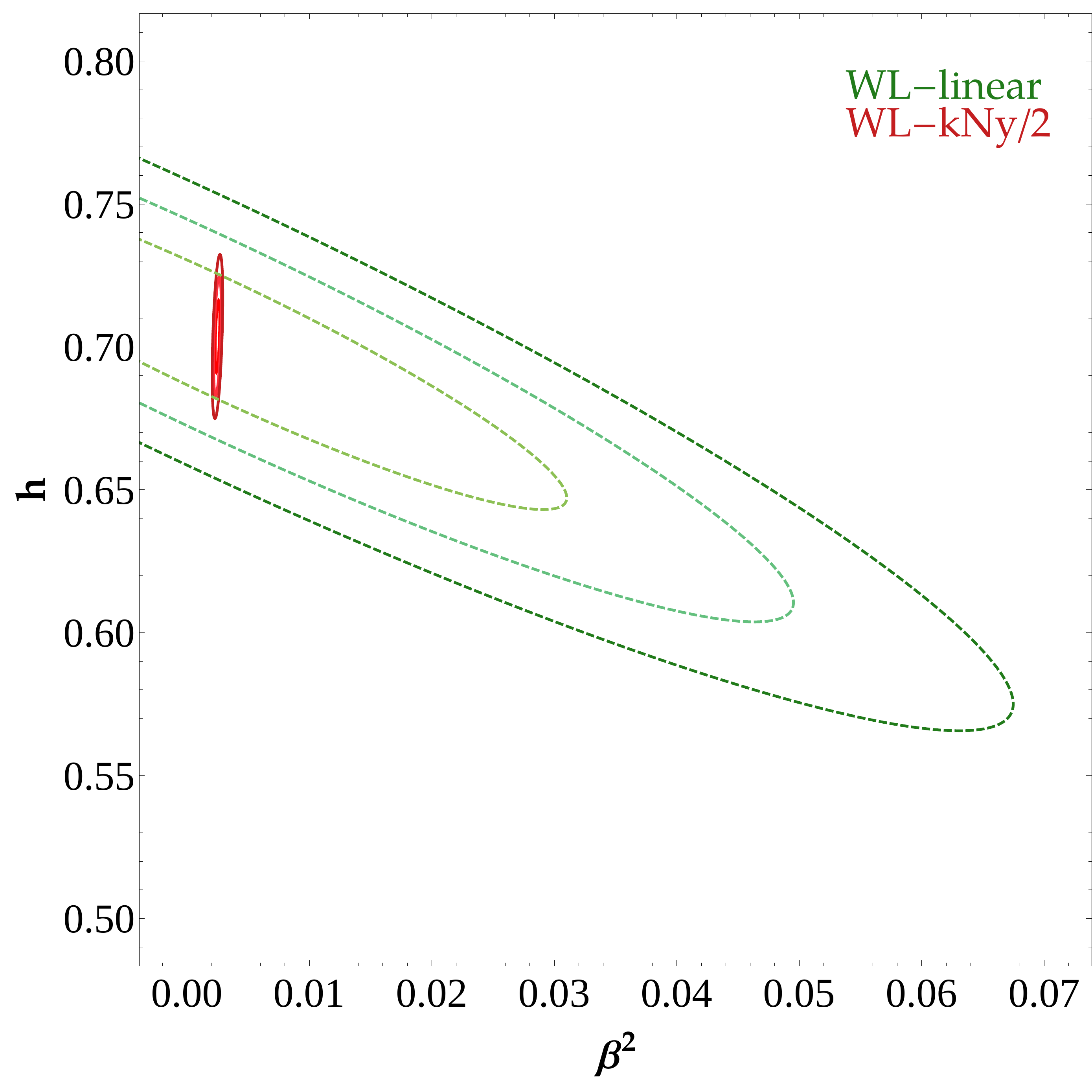}
\protect\protect\protect\caption{\label{fig:Contour-scale} Marginalized confidence contour regions
1,2,3-$\sigma$ for $\beta^{2}$ and $h$. The plots correspond to
a comparison between linear (green-dashed lines) and non-linear (red-solid
lines) scales for GC (left) and WL (right) using the scale cutoff
at $k_{Ny}/2$. }
\end{figure}

\subsection{Weak lensing}

Table \ref{tab:WL-specifications-1} show the specifications for a
weak lensing probe in an Euclid-like survey. The redshift bins are
chosen in such a way that they contain an equal number of galaxies
(equipopulated bins). The bins are then given by:

\begin{equation}
n_{i}(z)=\frac{1}{2}n(z)\left[\mbox{Erf}\left(\frac{\tilde{z}_{i+1}-z}{\sigma_{pz}\sqrt{2}}\right)-\mbox{Erf}\left(\frac{\tilde{z}_{i}-z}{\sigma_{pz}\sqrt{2}}\right)\right]
\end{equation}
where $\tilde{z}_{i}$ are the values of the bin intervals in the
range $z_{range}=0.5\leq z\leq3$ chosen such that for each interval
the integral over the galaxy distribution function $n(z)$ (eqn. \ref{eq:ngal dist})
is equal. The resulting peaks of the bins and their full width at
half maximum are specified in table \ref{tab:WL-zbins-specs}.

\begin{table}
\centering{}%
\begin{tabular}{|c|c|c|}
\hline 
\textbf{Parameter}  & \textbf{Value}  & \textbf{Description}\tabularnewline
\hline 
$f_{sky}$  & 0.364  & Observed sky fraction\tabularnewline
$\gamma_{int}$  & 0.22  & Intrinsic alignment\tabularnewline
$n_{\theta}$  & 30  & Galaxy density per $arcmin^{2}$\tabularnewline
$\sigma_{pz}$  & 0.05  & Photometric redshift error\tabularnewline
\hline 
\end{tabular}\protect\protect\caption{\label{tab:WL-specifications-1} Specifications for the Fisher Matrix
of an Euclid-like weak lensing survey.}
\end{table}

\begin{table}
\centering{}%
\begin{tabular}{|c|c|c|}
\hline 
\textbf{Parameter}  & \textbf{Value}  & \textbf{Description}\tabularnewline
\hline 
$\mathcal{N}$  & 6  & Number of redshift bins\tabularnewline
$z_{peak}$  & $\{0.59,0.75,0.90,1.06,1.28,1.57\}$  & $z$-position of peak of the bin\tabularnewline
$w_{z}$  & $\{0.22,0.23,0.25,0.27,0.32,0.51\}$  & full width at half maximum of the peak\tabularnewline
$\ell_{cut,\, k_{1}}$  & $\{1686,2070,2410,2753,3155,4344\}$  & cutoff in multipole $\ell$ at the center of each bin for $k_{max}=k_{Ny}/2$\tabularnewline
$\ell_{cut,\, k_{2}}$  & $\{3372,4141,4820,5506,6311,8689\}$  & cutoff in multipole $\ell$ at the center of each bin for $k_{max}=k_{Ny}$\tabularnewline
$z_{range}$  & $0.5\leq z\leq3$  & Total range in redshift of each bin\tabularnewline
\hline 
\end{tabular}\protect\protect\caption{\label{tab:WL-zbins-specs} Redshift bins specifications for an Euclid-like
weak lensing survey using equipopulated redshift bins in the range
$0\leq z\leq3$ and the corresponding values for the cutoff applied
in the multipoles $\ell$ at each redshift bin, for two different
cases: scales larger than $k_{1}=k_{Ny}/2$ and scales larger than
$k_{2}=k_{Ny}$.}
\end{table}

Analogously to the galaxy clustering case, we show in table \ref{tab:1-sigmaerrors-nl-WL-full}
the fully marginalised 1-$\sigma$ errors on the parameters $p_{i}$.
The sum over multipoles $\ell$ in eq. (\ref{eq:FisherSum-WL}) is
performed from $\ell_{min}=5$ up to a maximum of $\ell_{max}=20000$,
but as explained in section \ref{sub:Weak-Lensing}, we perform a
cutoff at each redshift bin, so that no scales in the non-linear power
spectrum beyond the half of the Nyquist (for our reference case) or
beyond the Nyquist frequency (for our second case) contribute to the
WL signal. The values of these cutoffs, $\ell_{cut}$ for the two
different cases $k_{1}=k_{Ny}/2$ and $k_{2}=k_{Ny}$ are listed in
table \ref{tab:WL-zbins-specs}. In contrast to the GC case, going
from $k_{Ny}/2$ to $k_{Ny}$ in a WL survey does bring a noticeable
improvement on the estimation of parameters.

\begin{table}
\centering{}%
\begin{tabular}{|c|c|c|c|c|c|c|}
\hline 
\textbf{Parameter}  & \textbf{$\text{\ensuremath{\beta}}^{2}$}  & $h$  & $10^{9}\mathcal{A}_{s}$  & $n_{s}$  & $\Omega_{b}$  & $\Omega_{c}$\tabularnewline
\hline 
fiducial  & 0.0025  & 0.7036  & 2.42  & 0.966  & 0.04503  & 0.2256\tabularnewline
\hline 
\textbf{$\mathbf{\boldsymbol{\ell}}_{\mathbf{cut,\, k_{Ny}/2}}$}  &  &  &  &  &  & \tabularnewline
\hline 
abs. error  & 0.000125  & 0.00835  & 0.112  & 0.0105  & 0.0032  & 0.0046\tabularnewline
relative errror & 5.0\%  & 1.2\%  & 4.6\%  & 1.1\%  & 7.1\%  & 2.0\%\tabularnewline
\hline 
\textbf{$\boldsymbol{\ell}_{\mathbf{cut,\, k_{Ny}}}$}  &  &  &  &  &  & \tabularnewline
\hline 
abs. error  & 0.000097  & 0.0068  & 0.058  & 0.0085  & 0.0022  & 0.0032\tabularnewline
relative error & 3.9\%  & 0.97\%  & 2.4\%  & 0.88\%  & 4.9\%  & 1.4\%\tabularnewline
\hline 
\end{tabular}\protect\protect\caption{\label{tab:1-sigmaerrors-nl-WL-full} 1-$\sigma$ fully marginalized
errors on the cosmological parameters of the model for a weak lensing
Fisher forecast. Two cases are presented, a redshift-dependent cutoff
$\ell_{cut,\: k_{Ny}/2}$ and a cutoff $\ell_{cut,\: k_{Ny}}$, corresponding
to cutting off the non-linear power spectrum at the half and at the
full Nyquist frequency (analogously to the GC case) as explained in
the main text. As opposed to the GC case, going into smaller scales
in a WL survey does bring a noticeable improvement. }
\end{table}

\begin{figure}[h]
\centering{}\includegraphics[width=0.7\textwidth]{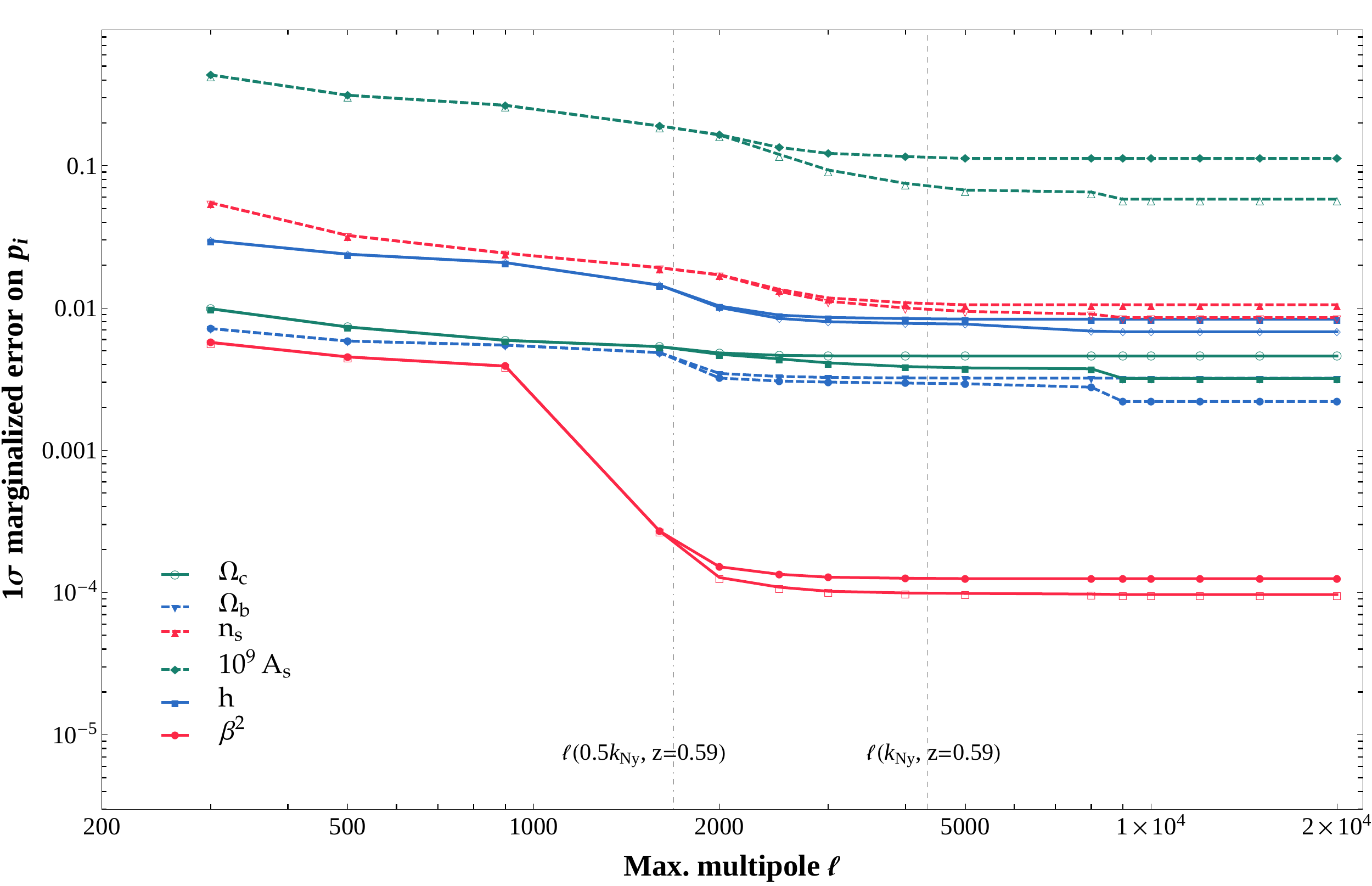}\protect\protect\protect\caption{\label{fig:WL-lmax-variation} Variation in the 1-$\sigma$ fully
marginalized errors for each cosmological parameter $p_{i}$ as a
function of the maximum multipole used in the weak lensing Fisher
forecast. Increasing the observed multipole range improves considerably
the determination of a parameter, especially on the coupling $\beta^{2}$
(red solid line) and the initial amplitude of perturbations $10^{9}A_{s}$
(dashed green line). The double lines corresponding to each parameter,
show how the error changes if a cut in the matter power spectrum $P(k)$
is performed at $k_{Ny}$ (lower line) instead of a cut at $k_{Ny}/2$
(upper line, respectively). The vertical dashed grey lines mark the
$\ell_{cut}$ at the peak of the first redshift bins for the cases
$\ell_{cut,\: k_{Ny}/2}$ and $\ell_{cut,\: k_{Ny}}$.}
\end{figure}

The dependence of the 1-$\sigma$ fully marginalized error on each
parameter $p_{i}$ with respect to $\ell_{max}$ is shown in figure
\ref{fig:WL-lmax-variation}. When using just linear power spectra
information, the error on the parameters does not improve if one increases
the scale $\ell_{max}$, while in the case where non-linear information
is used, increasing the maximum multipole $\ell_{max}$ improves considerably
the 1-$\sigma$ error on the parameters, especially on the coupling
$\beta^{2}$ and on the initial amplitude of scalar fluctuations $10^{9}A_{s}$.
This is due to the fact that the extra signal on the coupling coming
from the non-linear part of the power spectrum, the so called ``bump'',
greatly enhances the constraints on the parameter estimation. The
double lines corresponding to each parameter in figure \ref{fig:WL-lmax-variation},
show how the error estimation is changed if scales up to $k_{Ny}$
are included (lower line) compared to the upper line where only scales
up to $k_{Ny}/2$ contribute. At small $\ell$ both lines are on top
of each other and only start diverging at around $\ell=2000$, when
the extra amount of information contained in highly non-linear scales
starts becoming important. The most significant gains occur again
on the parameters $\beta^{2}$ and $10^{9}A_{s}$.

\subsection{Combined results}

\begin{figure}
\begin{centering}
\includegraphics[height=0.15\paperheight]{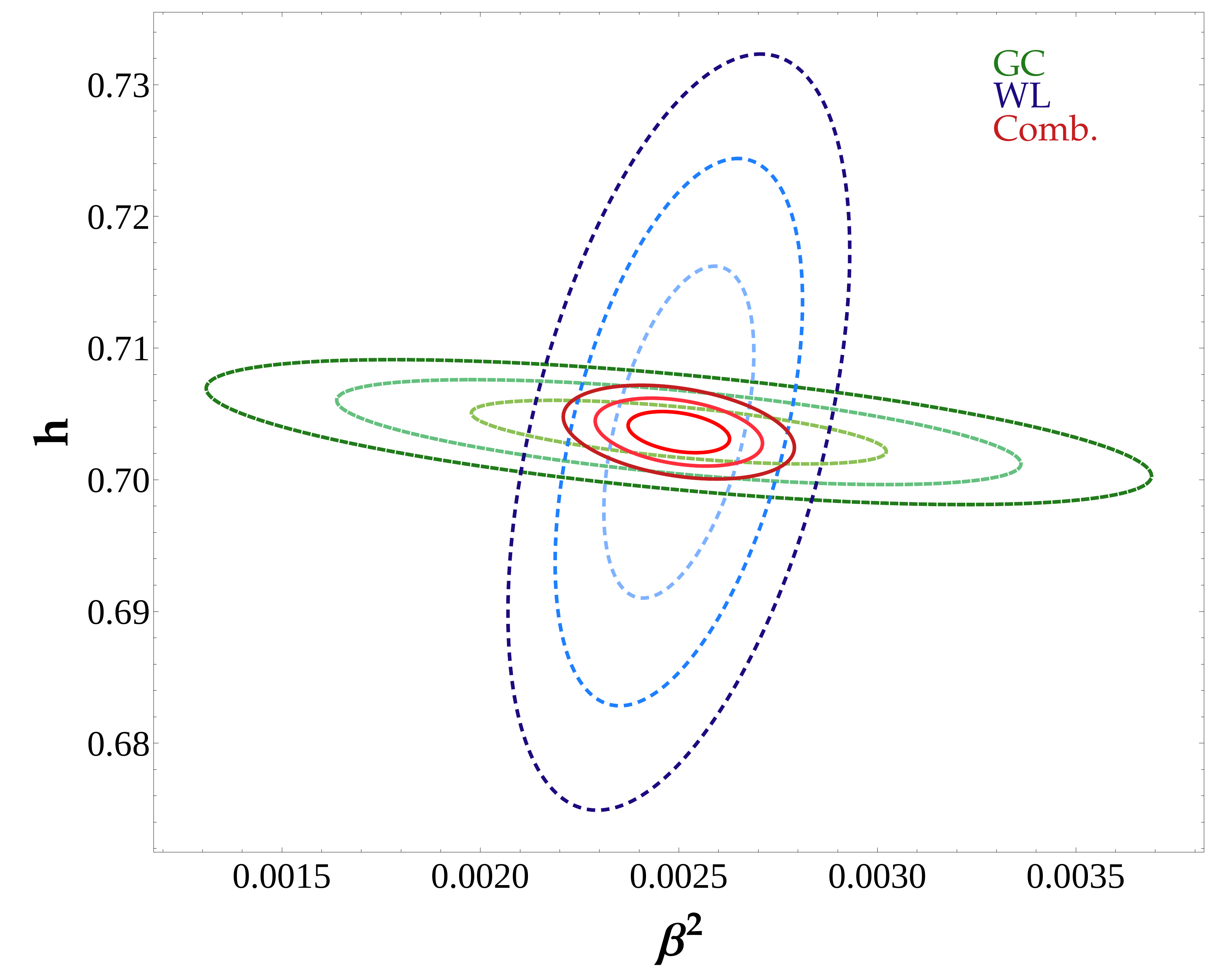}\hfill{}\includegraphics[height=0.15\paperheight]{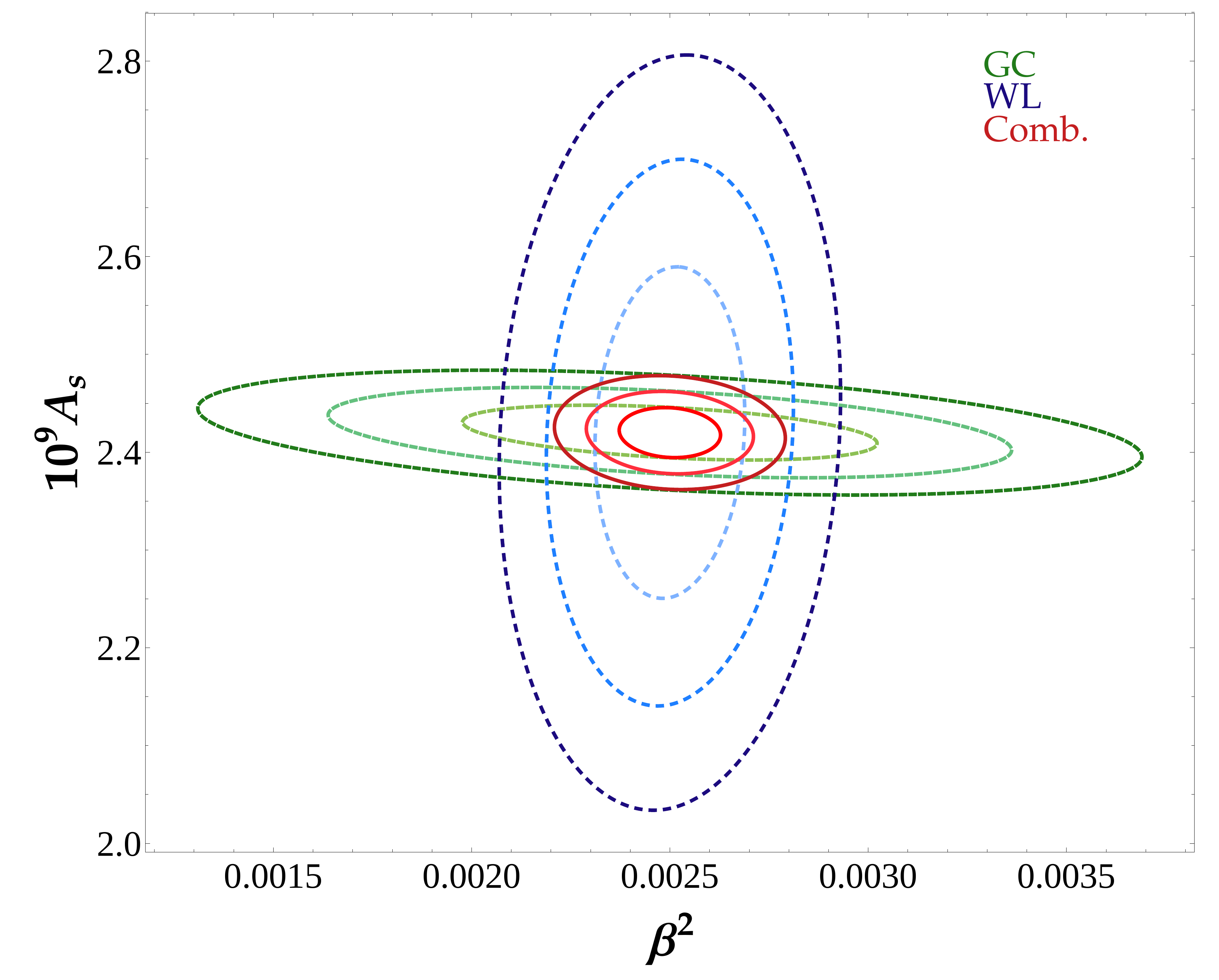}\hfill{}\includegraphics[height=0.15\paperheight]{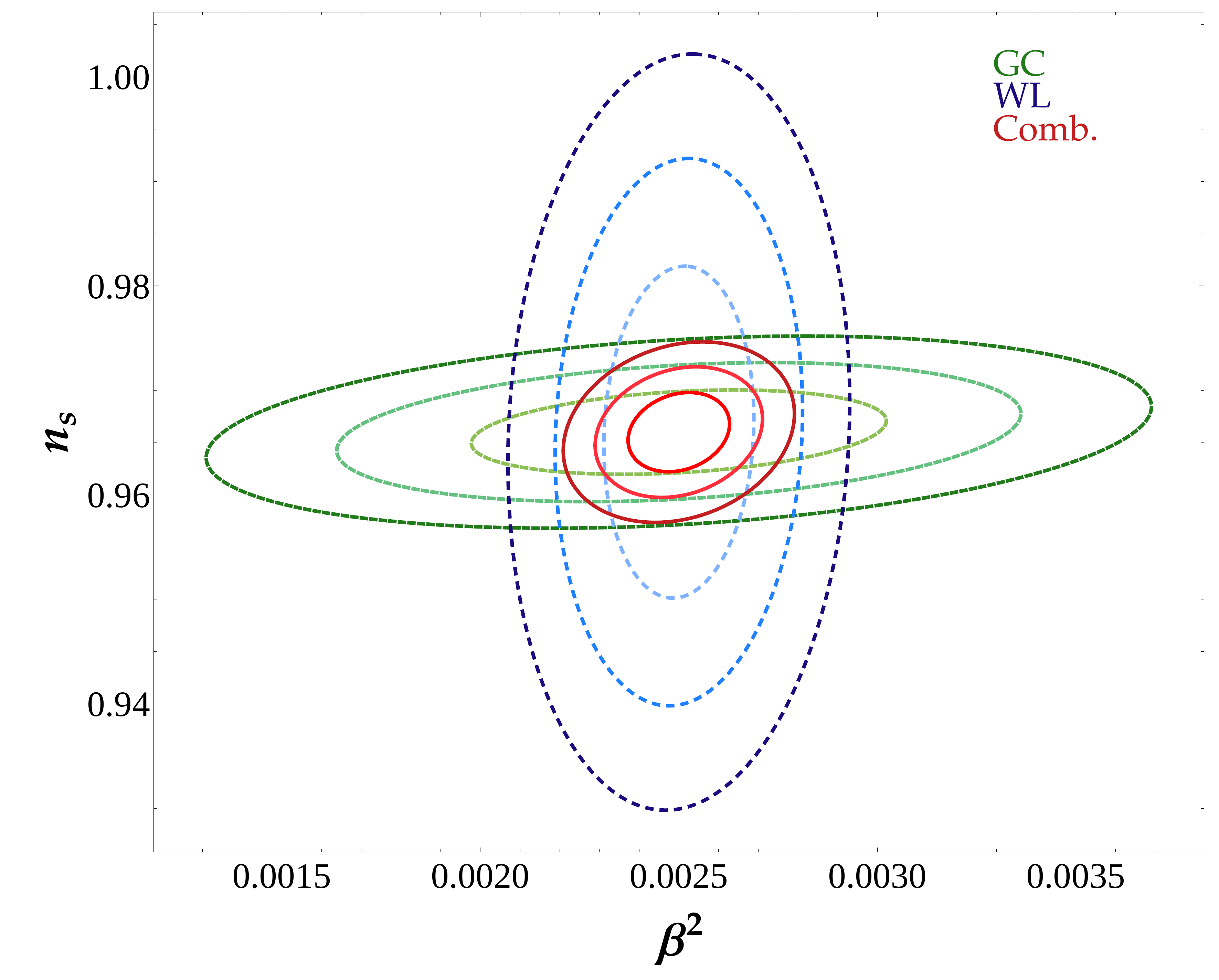} 
\par\end{centering}

\begin{centering}
\includegraphics[height=0.15\paperheight]{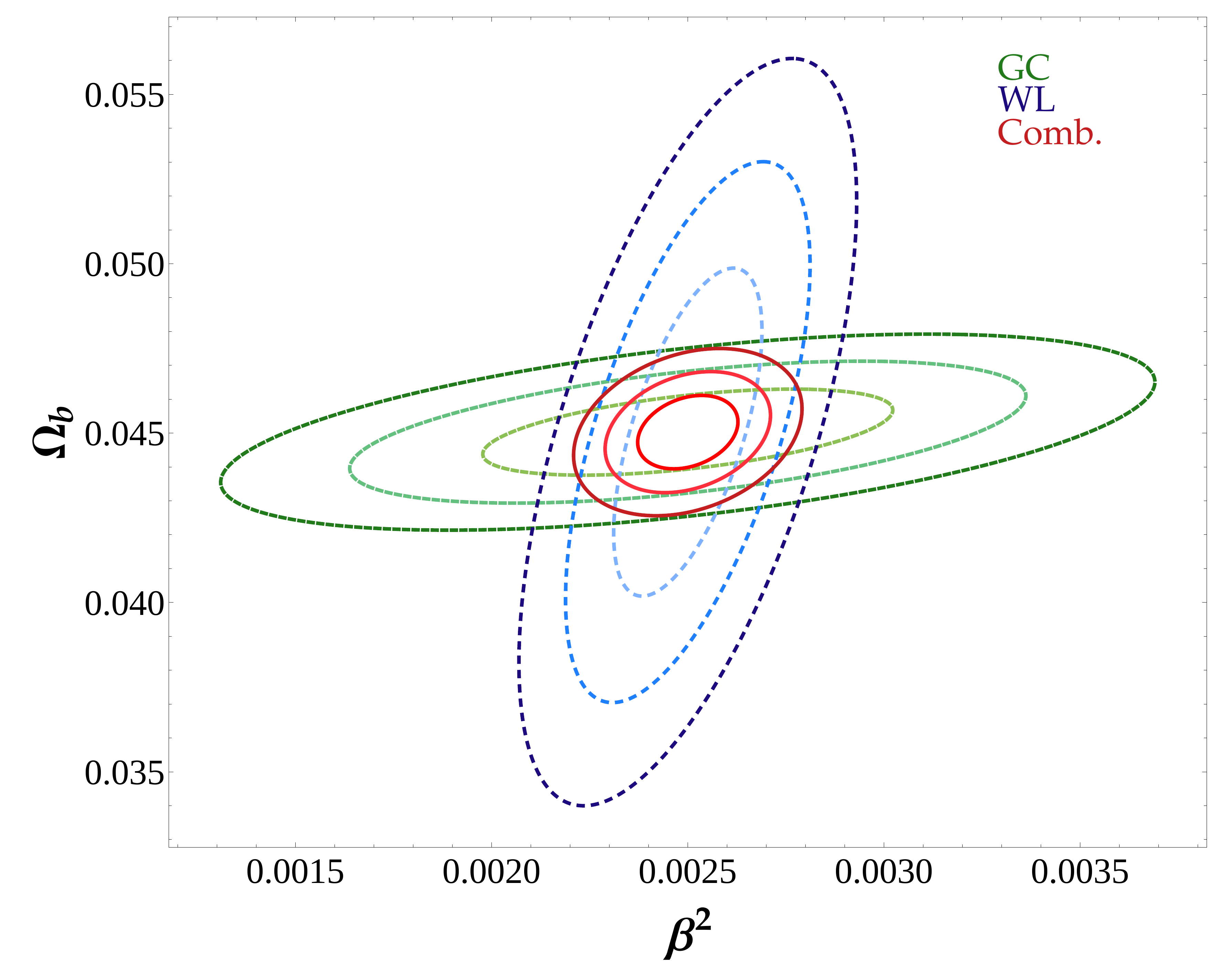}\hfill{}\includegraphics[height=0.15\paperheight]{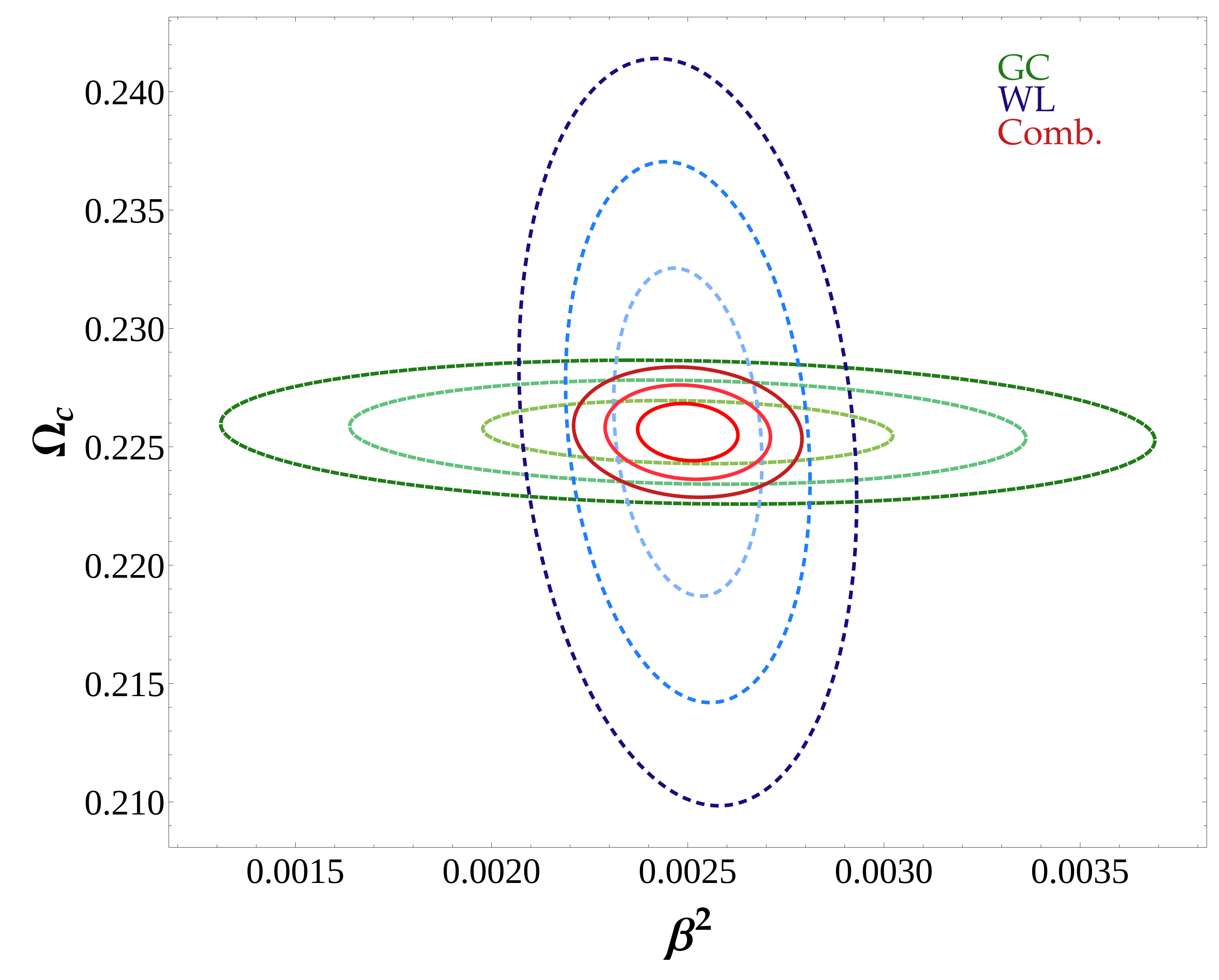}\hfill{}\includegraphics[height=0.15\paperheight]{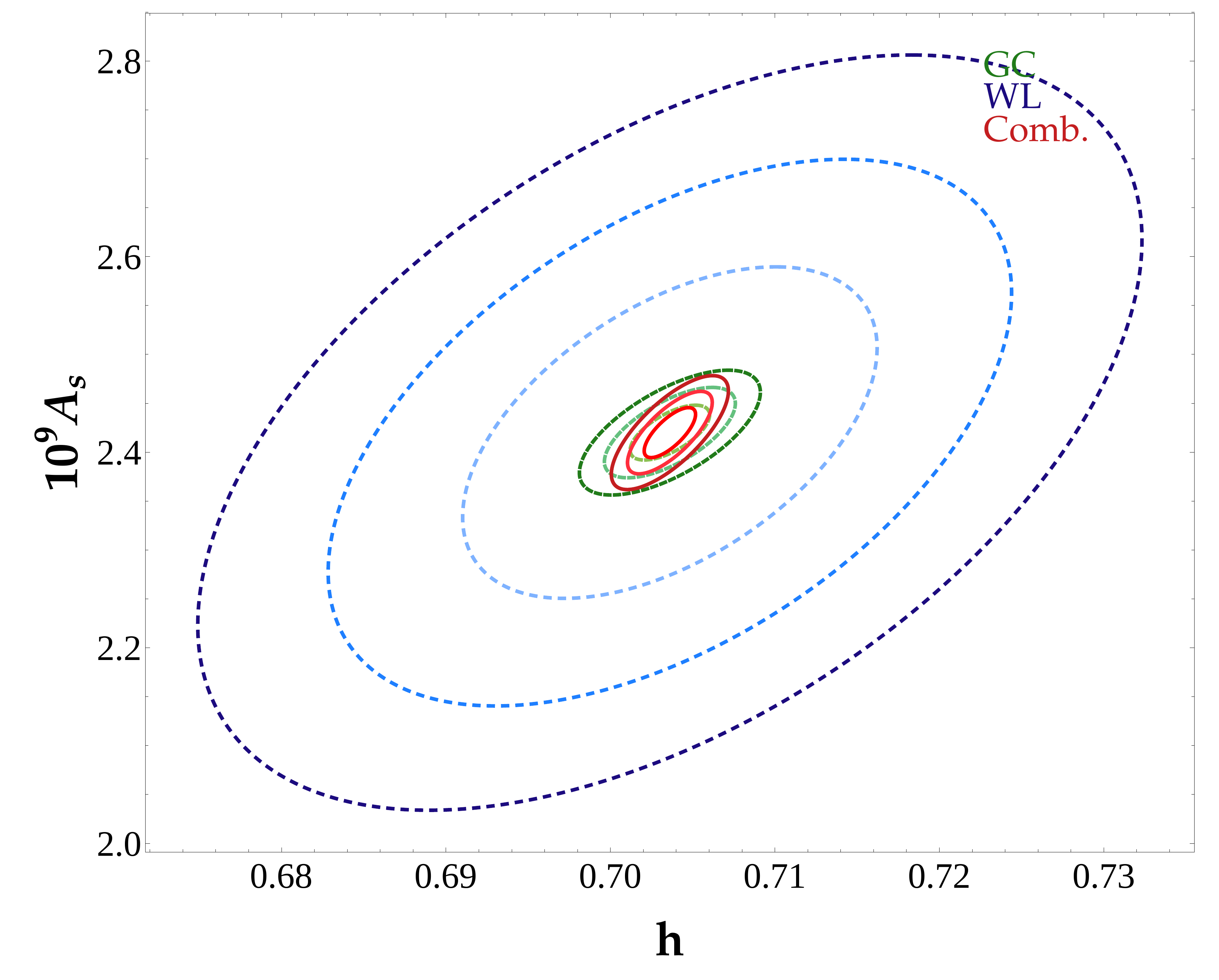} 
\par\end{centering}

\begin{centering}
\includegraphics[height=0.15\paperheight]{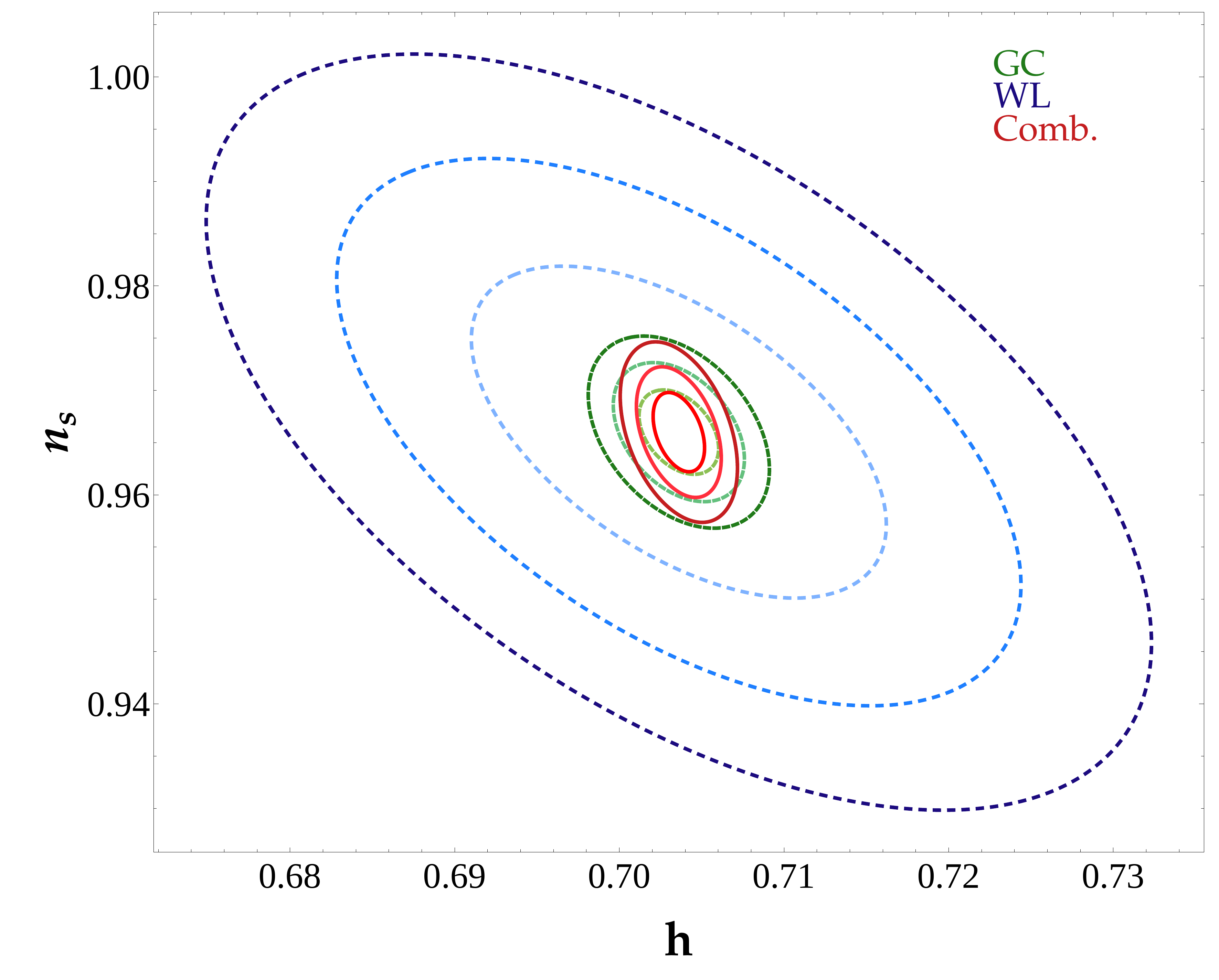}\hfill{}\includegraphics[height=0.15\paperheight]{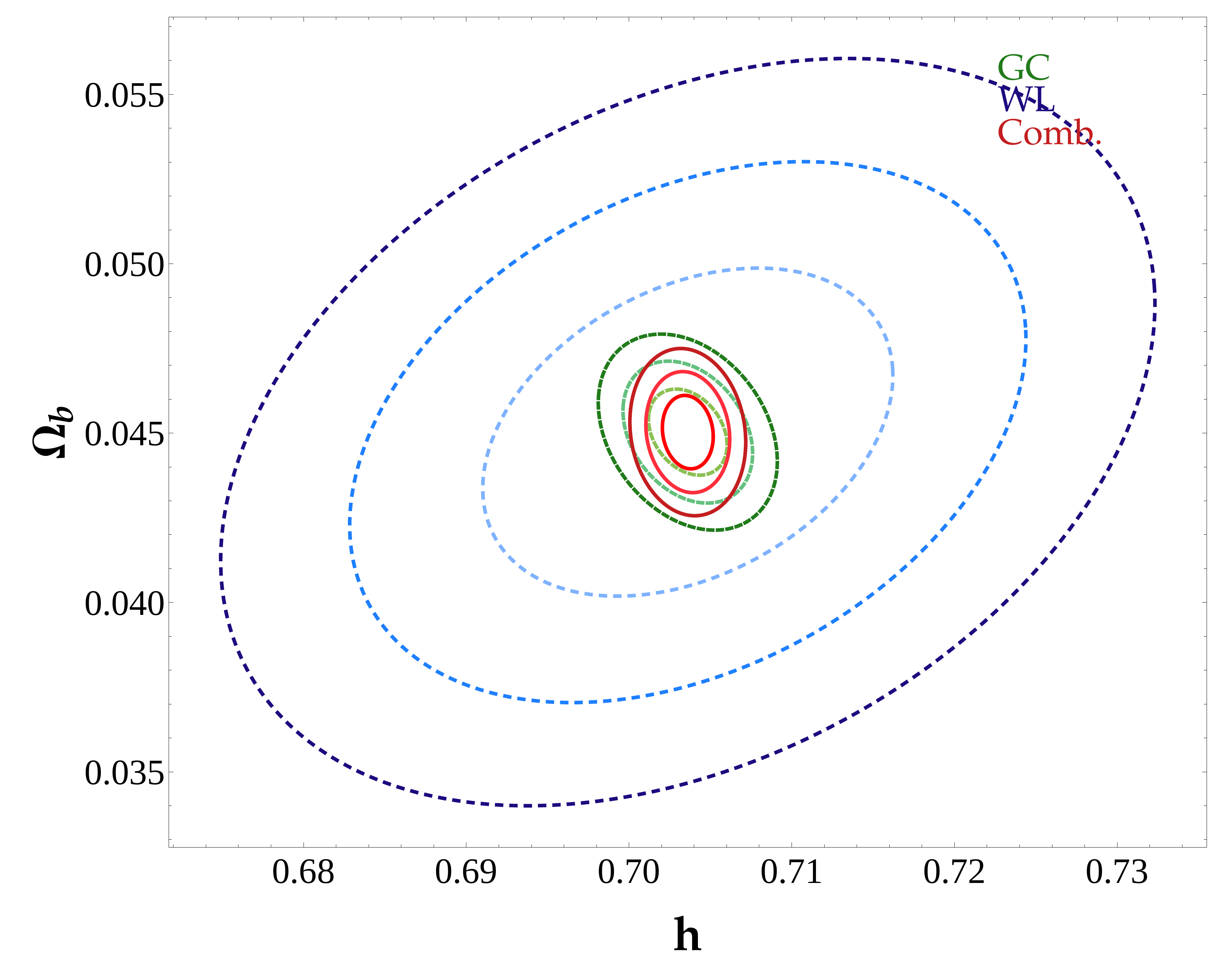}\hfill{}\includegraphics[height=0.15\paperheight]{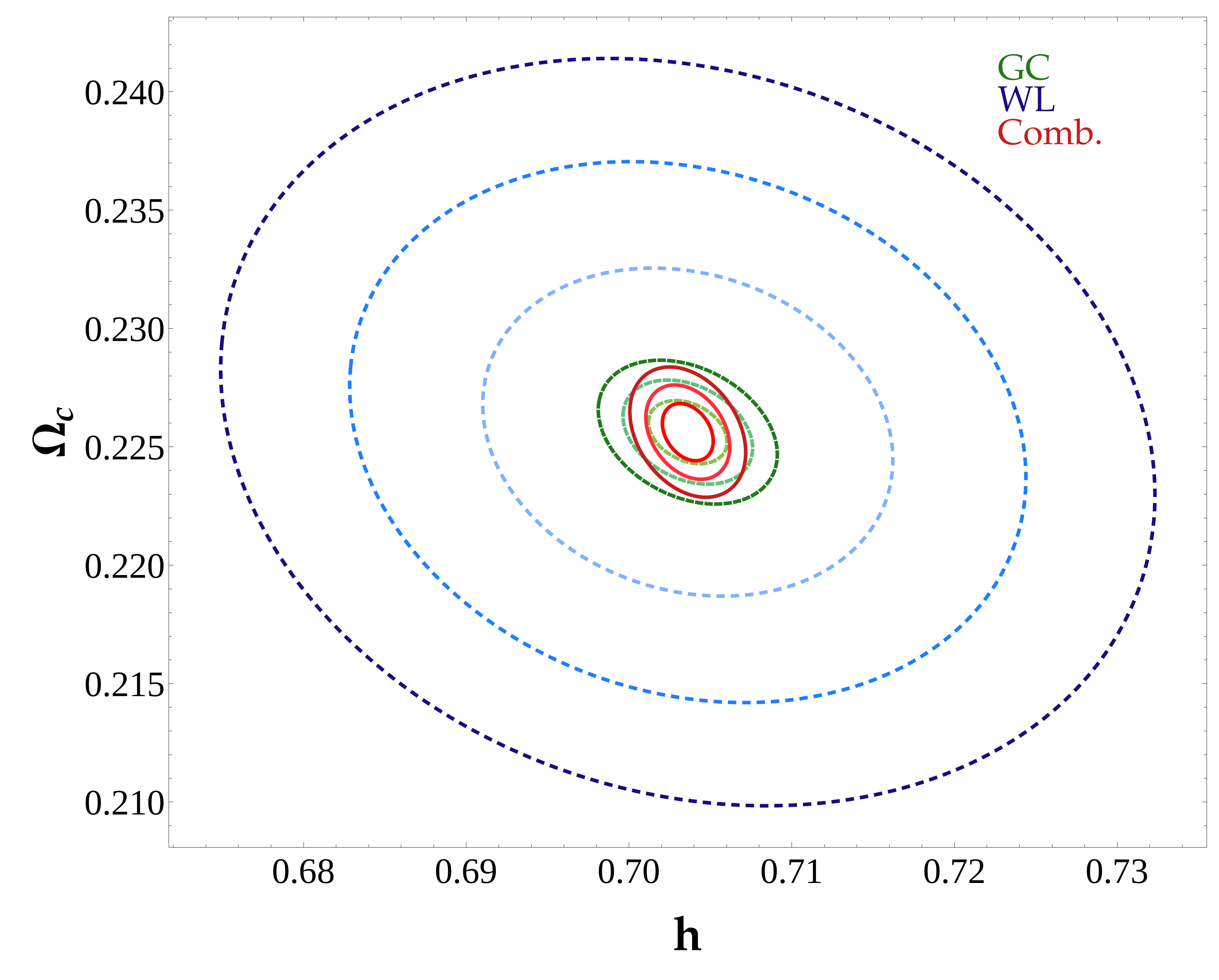} 
\par\end{centering}

\begin{centering}
\includegraphics[height=0.15\paperheight]{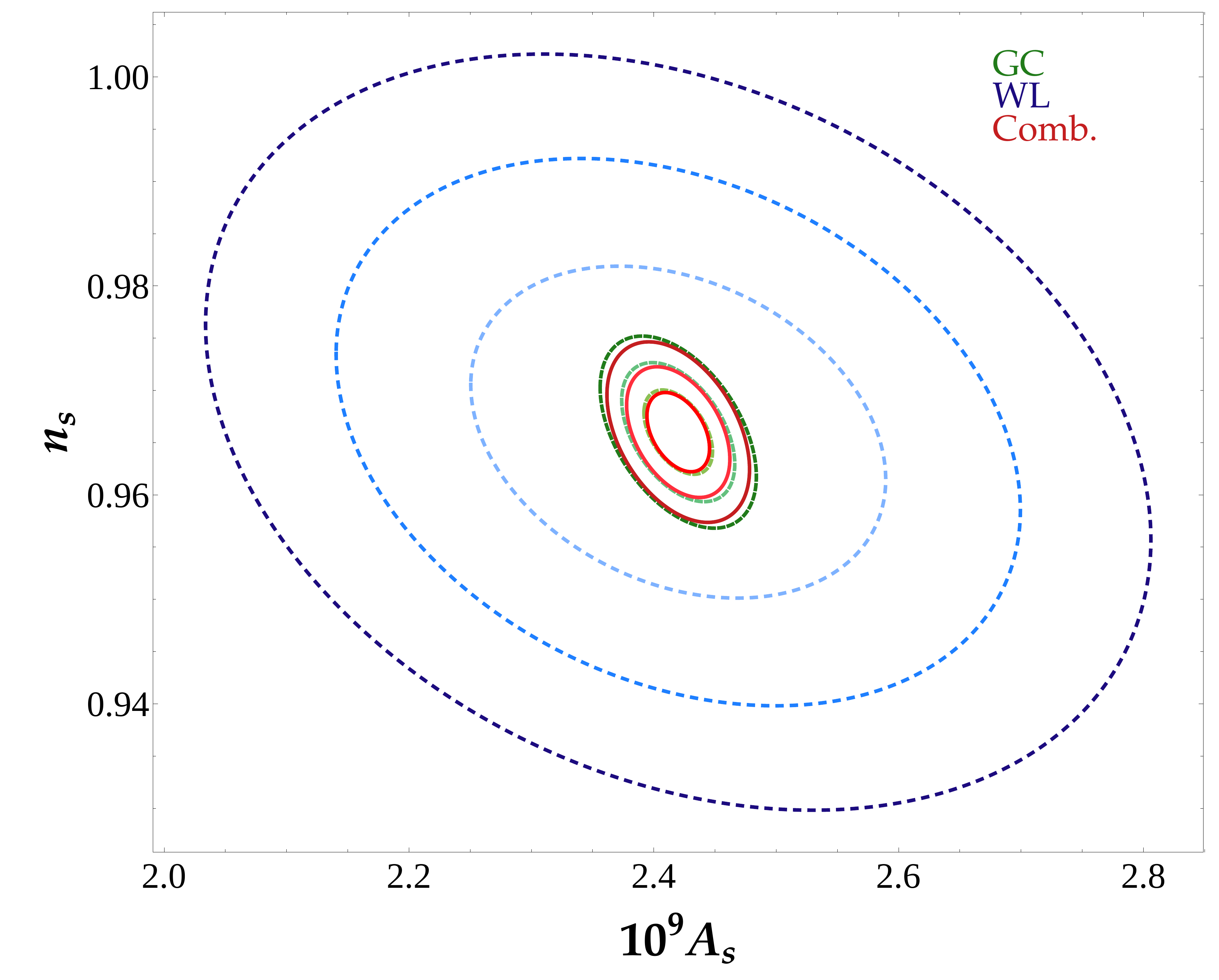}\hfill{}\includegraphics[height=0.15\paperheight]{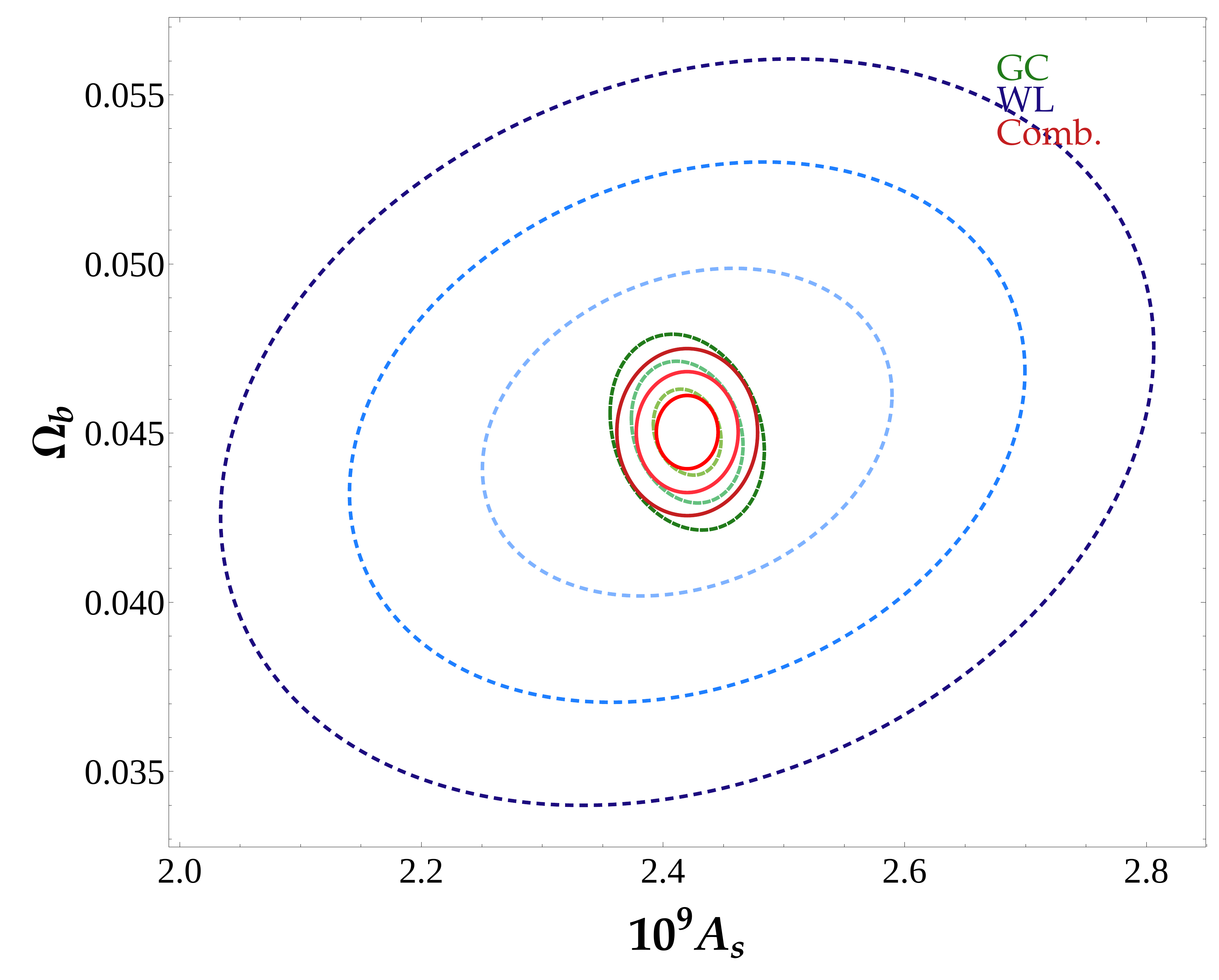}\hfill{}\includegraphics[height=0.15\paperheight]{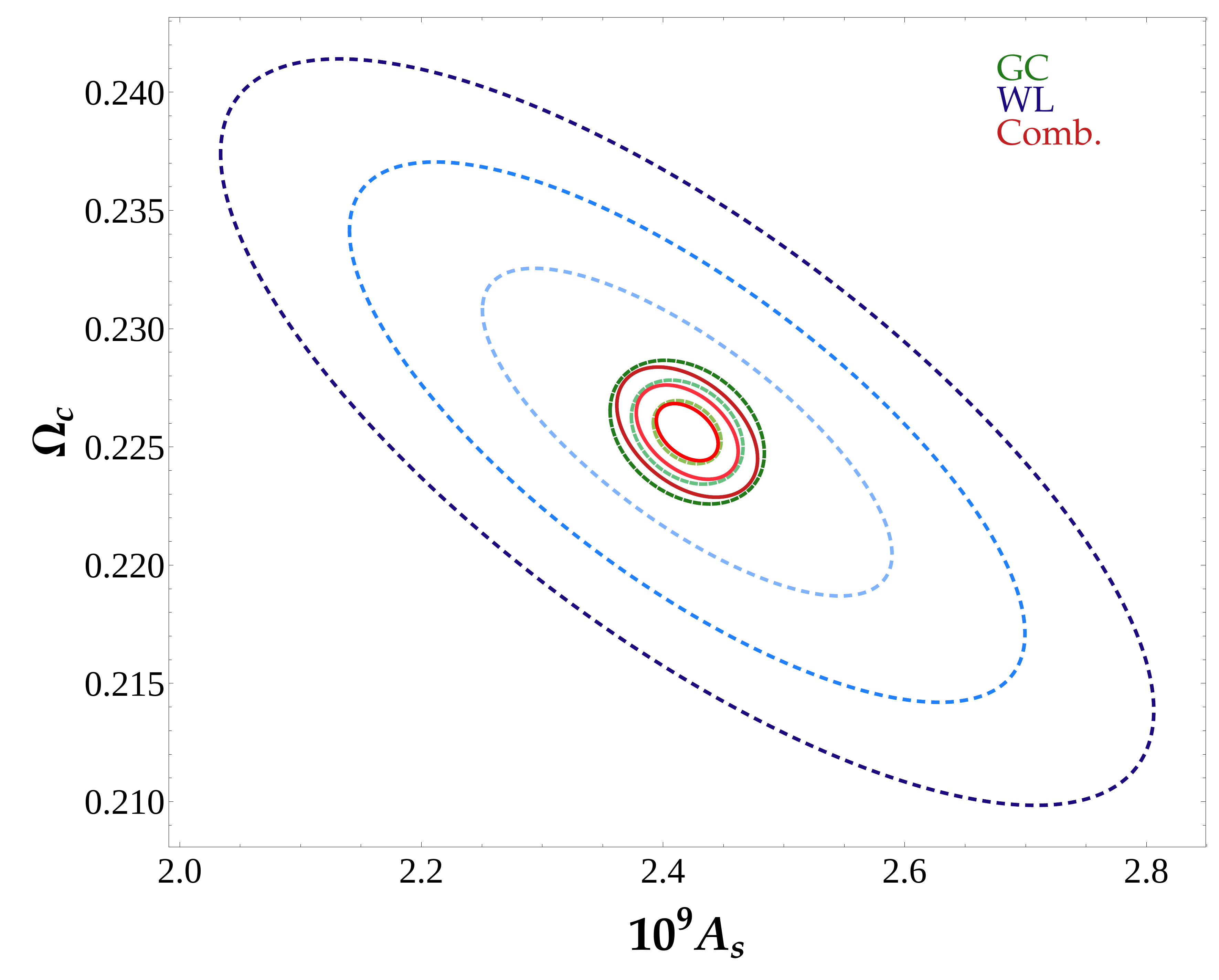} 
\par\end{centering}

\begin{centering}
\includegraphics[height=0.15\paperheight]{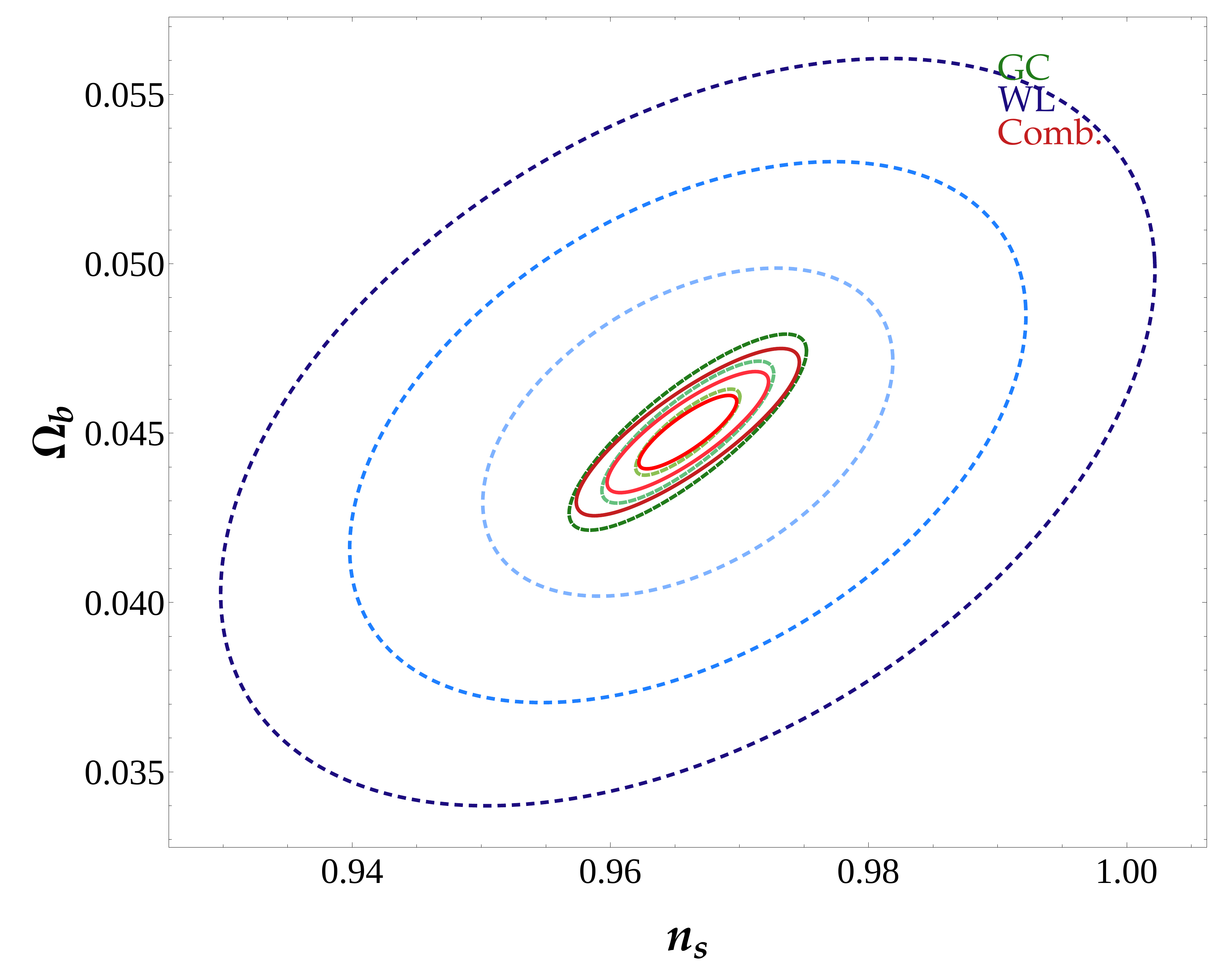}\hfill{}\includegraphics[height=0.15\paperheight]{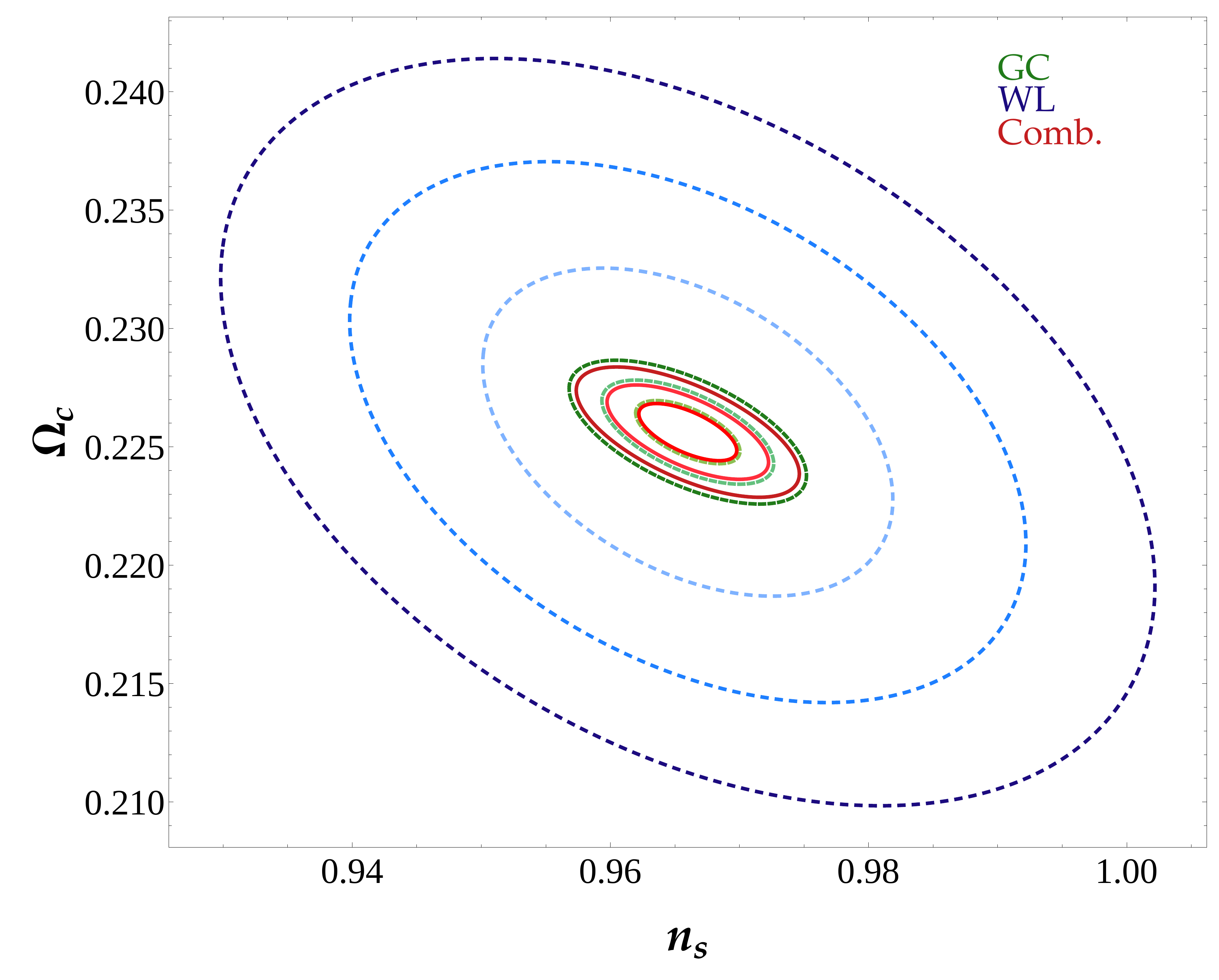}\hfill{}\includegraphics[height=0.15\paperheight]{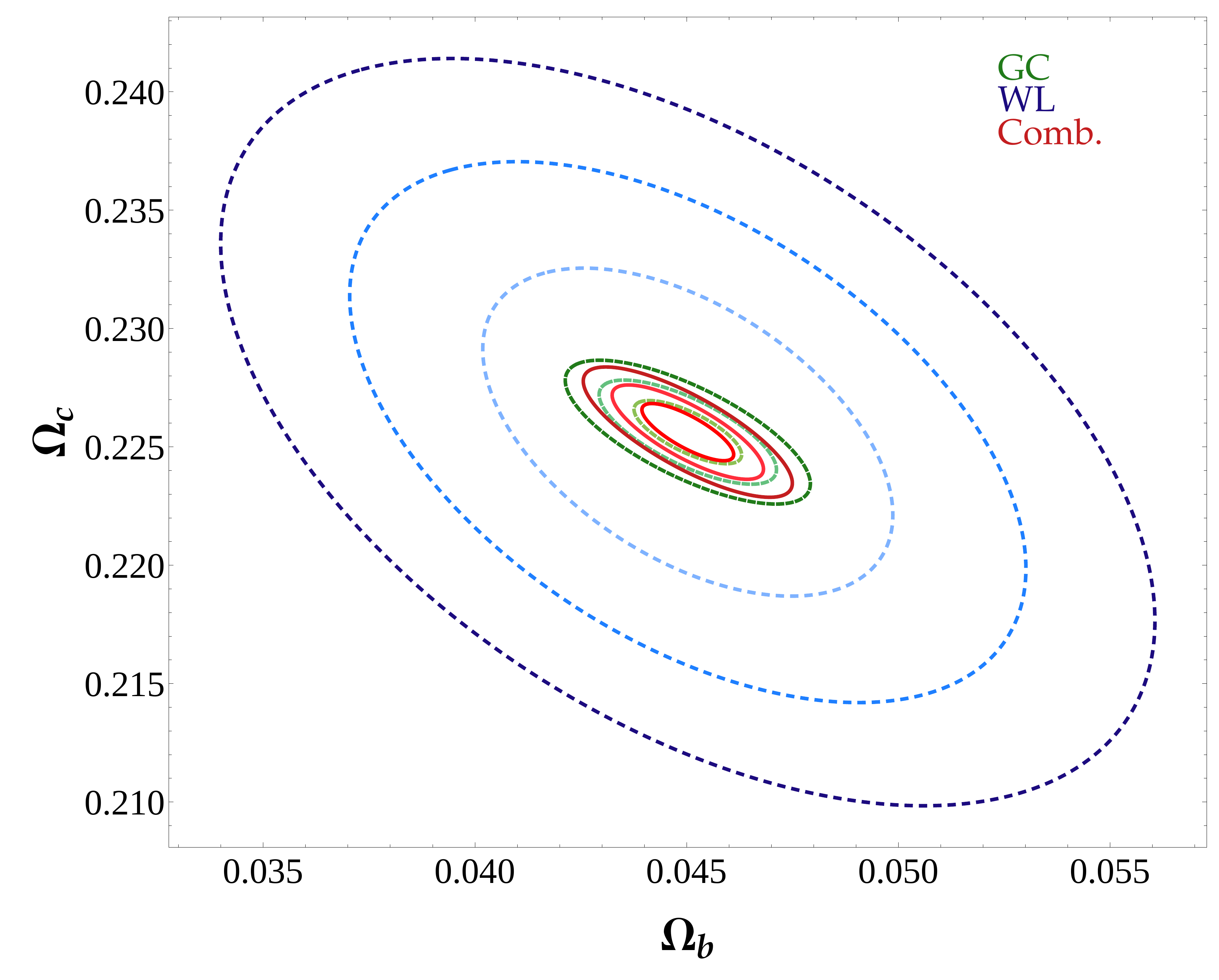} 
\par\end{centering}

\centering{}\protect\protect\protect\caption{\label{fig:Contour-regions} Marginalized confidence contour regions
(1,2,3-$\sigma$) for all cosmological parameters considered in this
model. The blue dashed lines correspond to the WL Fisher forecast,
while the green dashed lines correspond to the GC Fisher forecast
both in our reference case. The red solid lines correspond to the
combined Fisher matrix forecast. For combinations of the parameter
$\beta^{2}$, WL and GC have similar figures of merit, but different
orientations, while for other combinations of cosmological parameters,
the estimation is dominated by the GC Fisher matrix estimation.}
\end{figure}

In figure \ref{fig:Contour-regions} we show the 1-, 2- and 3-$\sigma$
confidence contours from the Fisher forecast for WL and GC. These
confidence regions for each pair of parameters are obtained after
marginalizing over all the other parameters. As it can be seen, some
degeneracies are broken when combining the confidence ellipses from
two different observations, for example in the case of the plane $\Omega_{b},\,\beta^{2}$.
Other parameter combinations, as $n_{s},\,\beta^{2}$, show the same
orientation of the ellipses for WL and GC, so that the combination
of both probes does not help to disentangle the degeneracies. While
GC constraints much better the usual parameters $\{h,\,10^{9}\mathcal{A}_{s},\, n_{s},\,\Omega_{c},\,\Omega_{b}\}$,
WL constrains the coupling parameter $\beta^{2}$ much better which
can be seen in the vertical orientation of the ellipses that correspond
to $\beta^{2}$. Therefore combining the observations on GC and WL,
as a future survey like Euclid will do, is a powerful way of constraining
degenerate parameters in cosmology.

In table \ref{tab:1-sigmaerrors-nl-WL-1-1}, we cite the 1-$\sigma$
fully-marginalised errors on the parameters $p_{i}$ for three different
cases: a) using only linear CDE power spectra computed from our modified
version of CAMB; b) applying a non-linear correction to these linear
CDE power spectra using the latest version of Halofit from \cite{takahashi_revising_2012},
which was designed for $\lcdm$-only; c) using the full coupled DE
non-linear power spectra computed with our fitting functions following
the procedure explained in section \ref{sub:Adding-non-linear-corrections}.

This shows the value of using the N-body-calibrated fitting functions
on the coupling $\beta^{2}$. Using the proper $\beta$-dependent
non-linear correction instead of the standard Halofit correction,
the constraints on $\beta^{2}$ improve by more than an order of magnitude
for WL and by a factor of order three for GC. This makes very clear
the importance of applying non-linear corrections that depend on the
parameter to be tested.

\begin{table}
\begin{centering}
\begin{tabular}{|c|c|c|c|c|c|c|}
\hline 
\textbf{Parameter}  & \textbf{$\text{\ensuremath{\beta}}^{2}$}  & $h$  & $10^{9}\mathcal{A}_{s}$  & $n_{s}$  & $\Omega_{b}$  & $\Omega_{c}$\tabularnewline
\hline 
fiducial  & 0.0025  & 0.7036  & 2.42  & 0.966  & 0.04503  & 0.2256\tabularnewline
\hline 
\textbf{WL: 1-$\mathbf{\sigma}$ abs. error, using:}  &  &  &  &  &  & \tabularnewline
\hline 
linear CDE  & 0.0189  & 0.040  & 0.221  & 0.0139  & 0.0062  & 0.0127\tabularnewline
linear CDE+Halofit  & 0.0184  & 0.044  & 0.256  & 0.0109  & 0.0066  & 0.0079\tabularnewline
non-linear CDE fitting functions  & 0.000125  & 0.00835  & 0.112  & 0.0105  & 0.0032  & 0.0046\tabularnewline
\hline 
\textbf{GC: 1-$\mathbf{\sigma}$ abs. error, using:}  &  &  &  &  &  & \tabularnewline
\hline 
linear CDE  & 0.0038  & 0.0097  & 0.117  & 0.0176  & 0.0021  & 0.0055\tabularnewline
linear CDE+Halofit  & 0.0011  & 0.0029  & 0.024  & 0.0023  & 0.0007  & 0.0006\tabularnewline
non-linear CDE fitting functions  & 0.00035  & 0.0016  & 0.018  & 0.0027  & 0.0008  & 0.0009\tabularnewline
\textbf{comb. WL+GC: 1-$\mathbf{\sigma}$ error, using:}  &  &  &  &  &  & \tabularnewline
\hline 
non-linear CDE fitting functions (abs.)  & 0.000084  & 0.0010  & 0.0169  & 0.00251  & 0.00072  & 0.00080 \tabularnewline
non-linear CDE fitting functions (rel.)  & 3.4\%  & 0.16\%  & 0.7\%  & 0.26\%  & 1.6\%  & 0.35\%\tabularnewline
\hline 
\end{tabular}
\par\end{centering}

\protect\protect\caption{\label{tab:1-sigmaerrors-nl-WL-1-1} 1-$\sigma$ fully marginalized
errors on the cosmological parameters for WL, GC and the combined
Fisher matrix WL+GC, using three different power spectra. \textbf{Linear
CDE}: Using only information from the linear power spectrum for the
CDE model up to a scale of $k=0.1h/\mbox{Mpc}$. \textbf{Linear CDE+Halofit:
}Using the linear power spectrum for CDE plus a non-linear correction
using the latest Halofit from \cite{takahashi_revising_2012}. \textbf{Non-linear
CDE fitting functions: }Using the fully non-linear power spectra for
CDE obtained from the fitting functions and the emulator as explained
in \ref{sec:Fitting-functions}, which we regard as the most reliable
description of the model in this range of scales. In all these cases
we are using our reference Fisher forecasts corresponding to the cutoff
at $k_{Ny}/2$. }
\end{table}

\newpage{}

\section{Conclusions}

The goal of this paper is to exploit the cosmological information
contained in the non-linear regime in order to improve parameter estimation
from future large-scale observations. The main obstacle along this
road is that we have accurate non-linear corrections for the matter
power spectrum only for $\Lambda$CDM and a few other relatively simple
variants, but not for the large variety of modifed gravity models
that have been proposed in recent years.

The first part of this paper has then been devoted to the task of
finding corrections to the linear power spectrum in the range of $k\approx0.1-1h/$Mpc
for a selected class of modified gravity models, namely coupled dark
energy. This model is indeed one of the simplest possible extensions
of Einstein's gravity and depends entirely on a single parameter,
the coupling constant $\beta$ (in addition to the standard ones).
Employing the CoDECS suite of simulations \citep{baldi_codecs_2012}
we build different fitting function models, such that when multiplied
by the $\Lambda$CDM non-linear power spectrum (we use the estimator
provided in ref. \citep{heitmann_coyote_2014} ) they reproduce the
N-body results to an accuracy of 1\%, for scales $k$ between 0.1
and 5 $h/\mbox{Mpc}$ and a range in $z$, between 0 and 1.8. To achieve
this accuracy in the fitting functions we need to perform a careful
extraction and interpolation of the power spectrum from the simulation
mesh.

The accurate fitting functions have been then employed to extend the
regime of validity of the forecasts for future experiments. We focused
on a Euclid-like survey that includes weak lensing and redshift-space
distortions (galaxy clustering) and predicted the constraints in the
cosmological parameters, with particular emphasis on the dark matter-dark
energy coupling $\beta$. We find that $\beta$ is better constrained
by weak lensing than by galaxy clustering (contrary to all the other
standard parameters). We find that the extension into non-linear scales
improves the constraints by more than an order of magnitude compared
to previous results using only linear power spectra, but also by more
than an order of magnitude in WL and a factor of three in GC compared
to using a wrong $\lcdm$ Halofit non-linear correction. We also show
that using the wrong non-linear power spectrum, can bias systematically
the estimation of errors on the cosmological parameters, yielding
systematic errors of the same order of magnitude as the statistical
ones. This makes very clear that it is important to include the proper
non-linear corrections to the parameter to be tested, especially for
models beyond $\lcdm$ in which the small-scale gravitational dynamics
are modified.

To make our forecast more realistic, we take into account all known
sources of error entering the estimation of the power spectra and
the fitting functions in the way of a reduced effective number density
of galaxies and then perform a conservative cut of the power spectrum
at half of the simulation Nyquist frequency, to avoid other sources
of unknown numerical noise affecting the results. In the case of GC
we include also a first approximation to the correction to redshift
space distortions, caused by peculiar pairwise velocities at non-linear
scales. We find that a space probe like Euclid will be able to constrain
the coupling parameter $\beta^{2}$ around the fiducial value $0.0025$
at 1-$\sigma$ with a relative accuracy of 14\% when using weak lensing
alone, 5\% when using only galaxy clustering and at 3.4\% when combining
both probes.

It is interesting to note that the most stringest constraint we obtain
amounts to $\Delta\beta^{2}\approx8\cdot10^{-5}$; this level of precision
on the dark matter-dark energy coupling is not far from the current
best limits reached with Solar System observations on a coupling to
baryons \cite{Agashe:2014kda}, which can be translated in our notation
as $\beta^{2}\le2\cdot10^{-5}$ at 1-$\sigma$. 
\begin{acknowledgments}
This research has been supported by DFG through the grant TRR33 ``The
Dark Universe''. V.P. acknowledges support from the Heidelberg Graduate
School for Fundamental Physics. S.C. acknowledges support from the
Heidelberg Graduate School for Fundamental Physics and a previous
support by the MICIT (Costa Rica). We thank E. Majerotto, D. Sapone,
S. Camera for useful discussions. 
\end{acknowledgments}

\bibliographystyle{unsrtnat}
\bibliography{paper-cited-new}

\clearpage{}

\rule[0.5ex]{1\columnwidth}{1pt}

\appendix

\section{Parameters of the fitting models M2 and M7\label{sec:AppendixA}}

Model M2 and M7 have each 5 coefficients, which are third order polynomial
function of $\beta$ and $z$:

\begin{align}
a_{i}\,(\beta,\, z) & =q_{i1}+q_{i2}\beta+q_{i3}\, z+q_{i4}\beta z\nonumber \\
 & +q_{i5}\beta^{2}+q_{i6}\, z^{2}+q_{i7}\, z^{3}+q_{i8}\,\beta^{3}\label{eq:a-params-funcs}
\end{align}

Since we have 5 coefficients and 8 free parameters for each coefficient,
we have a total of 40 free parameters. They can be represented in
a matrix $q_{ij}$ by: $A_{i}=q_{ij}B_{j}$, where the vector $A_{i}=a_{i}(\beta,\, z)$
is the vector of coefficients and $B_{j}=(1,\,\beta,\, z,\,\beta z,\,\beta^{2},\, z^{2},\,\beta^{3},\, z^{3})$
the vector of independent variables and the indices are defined in
the ranges: $i=1,...,5,\; j=1,...,8$. The corresponding matrix $q_{ij}$
for model M7 is:

\begin{minipage}[t]{1\textwidth}%
{\small{}{} 
\begin{equation}
q_{ij}(M7)=\begin{pmatrix}0.00689754 & -0.200001 & -0.0112943 & 0.157222 & -1.40171 & 0.0129681 & -3.99345 & -0.00629434\\
-0.0706221 & 1.45187 & 0.236402 & -1.26467 & 9.13266 & -0.300832 & 9.81754 & 0.120649\\
0.069102 & -1.44396 & -0.24268 & 1.26135 & -9.23052 & 0.311995 & -12.1568 & -0.12561\\
1.19797 & 0.632936 & -1.74767 & 0.920634 & -8.47056 & 1.46434 & 42.7015 & -0.471948\\
0.684838 & 1.15722 & -2.15405 & 3.93688 & -5.29011 & 3.42683 & -43.9295 & -1.18691
\end{pmatrix}\label{eq:qijM7}
\end{equation}
}%
\end{minipage}

and the coefficient matrix for model M2 is:

\begin{minipage}[t]{1\textwidth}%
{\small{}{} 
\begin{equation}
q_{ij}(M2)=\begin{pmatrix}0.00301472 & -0.0961879 & -0.00897697 & 0.0389941 & -0.860642 & 0.0131329 & -4.73786 & -0.00552858\\
-0.028584 & 0.615003 & 0.111671 & -0.371008 & 4.94379 & -0.138279 & 20.3258 & 0.0512362\\
0.0132728 & -0.345125 & -0.0701132 & 0.115389 & -3.11777 & 0.102871 & -16.9949 & -0.0408223\\
2.91757 & 1.03545 & -5.16595 & 1.69976 & -22.1649 & 4.33453 & 120.513 & -1.35074\\
0.867566 & 2.92363 & -1.14174 & 3.4356 & -2.96193 & 1.91413 & -113.348 & -0.684269
\end{pmatrix}\label{eq:qijM2}
\end{equation}
}%
\end{minipage}

Applying a singular value decomposition on both matrices, shows that
they have a very similar structure, therefore since the sigmoidal
functions M2 and M7 have a related functional form, the matrices describe
roughly the same information. 
\end{document}